\documentclass{aa}
\usepackage{amsmath}
\usepackage{txfonts}
\usepackage{natbib}
\usepackage{graphicx}
\usepackage{multirow} 
\graphicspath{{./YYGrid-Fig/}{./SODSPH-Fig/}{./SEDOVOFF-Fig/}
              {./POT-Fig/}{./RT-Fig/}{./POLYTROPE-Fig/}}
\bibpunct{(}{)}{;}{a}{}{,}

\def\ie{{\it i.e.,}\,}
\def\eg{{\it e.g.,}\,}
\title{An axis-free overset grid in spherical polar coordinates for 
       simulating 3D self-gravitating flows}
\abstract{} 
{Three dimensional explicit hydrodynamic codes based on spherical polar
coordinates using a single spherical polar grid suffer from a severe
restriction of the time step size due to the convergence of grid lines
near the poles of the coordinate system.  More importantly, numerical
artifacts are encountered at the symmetry axis of the grid where
boundary conditions have to be imposed that flaw the flow near the
axis.  The first problem can be eased and the second one avoided by
applying an overlapping grid technique. }
{A type of overlapping grid in spherical coordinates is adopted. This
so called ``Yin-Yang'' grid is a two-patch overset grid proposed by
Kageyama and Sato for geophysical simulations. Its two grid patches
contain only the low-latitude regions of the usual spherical polar
grid and are combined together in a simple manner.  This property of
the Yin-Yang grid greatly simplifies its implementation into a 3D code
already employing spherical polar coordinates. It further allows for a
much larger time step in 3D simulations using high angular resolution
($\la 1^\circ$) than that required in 3D simulations using a regular
spherical grid with the same angular resolution. }
{The Yin-Yang grid is successfully implemented into a 3D version of the
explicit Eulerian grid-based code PROMETHEUS including
self-gravity. The modified code successfully passed several standard
hydrodynamic tests producing results which are in very good agreement
with analytic solutions. Moreover, the solutions obtained with the
Yin-Yang grid exhibit no peculiar behaviour at the boundary between
the two grid patches. The code has also been successfully used to
model astrophysically relevant situations, namely equilibrium
polytropes, a Taylor-Sedov explosion, and Rayleigh-Taylor
instabilities. According to our results, the usage of the Yin-Yang
grid greatly enhances the suitability and efficiency of 3D explicit
Eulerian codes based on spherical polar coordinates for astrophysical
flows. }{}
\author{Annop Wongwathanarat \and Nicolay J. Hammer
\thanks{{\it present address:} Max-Planck-Institut f\"ur Plasmaphysik,
                               Boltz\-mann\-stra{\ss}e 2, D-85748 Garching}
        \and Ewald M\"{u}ller}
\institute{Max-Planck Institut f\"{u}r Astrophysik,
  Karl-Schwarzschild-Stra\ss e 1, D-85740 Garching, Germany}

\begin{document}
\maketitle

\section{Introduction}
Three dimensional hydrodynamic simulations employing a single
spherical polar grid are computationally expensive because of the
convergence of grid lines towards the north and south pole. The
converging grid lines imply a severe restriction of the time step size
for any hydrodynamic code using explicit time discretization due to
the CFL condition. This so-called ``pole problem'' bothers
astrophysicists when simulating self-gravitating flow in three
dimensions (e.g., convection in stars, or stellar explosions) where
the spherical coordinate system is often preferable. In particular,
simulations of core-collapse supernovae are a problem with which
astrophysicists have been struggling. While observations show clear
evidence of asymmetric (3D) complex structures in supernova ejecta,
numerical simulations, in most cases, are carried out only in two
spatial dimensions assuming axisymmetry \citep[e.g.,][]{2D2,2D1,2D3}.
Three dimensional core-collapse supernova simulations are rare
\citep[e.g.,][]{janka_etal_05, mezzacappa, scheck, iwakami}. In
addition to the severe restriction of the time step size, boundary
conditions that have to be imposed at the symmetry axis $\theta \in
[0, \pi]$ flaw the simulations near the axis by causing undesired
numerical artifacts in 2D axisymmetric simulations, as e.g., jet-like
flow features \citep{jet-like}. In 3D simulations, the axis represents
a coordinate singularity that almost unavoidably will leave its mark
on the flow near or across the axis.

There have been attempts to construct a new type of grid which is able
to ease the pole problem. However, it is not possible to construct a
single grid patch that can cover the entire surface of a sphere, is
orthogonal, and at the same time does not contain any coordinate
singularity except at the origin. Therefore, multi-patch grid and
overlapping (or overset) grid approaches are employed. They are widely
used in the field of computational fluid dynamics where complex grid
structures are common. For flows possessing an approximate global
spherical symmetry, the ``cubed sphere'' grid \citep{cubed-sphere} has
been developed and is currently applied to several astrophysical
problems \citep[e.g.]{cubed-app,cubed-app4,cubed-app3,cubed-app2}. 
It is an overset grid consisting of six
identical patches covering a solid angle of $4\pi$ steradians. The
``Yin-Yang'' grid has the latter property, too, but up to now it has
not been used in astrophysical applications.

The Yin-Yang grid was introduced by \citet{basicyy}. It consists of
two overlapping grid patches named ``Yin'' and ``Yang'' grid.  In
comparison with other types of overset grids in spherical geometry,
the Yin-Yang grid geometry is simple, as both the Yin and the Yang
grid consist of a part of a usual spherical polar grid. The
transformation of coordinates and vector components between the two
patches is straightforward and symmetric, thus allowing for an easy
and straightforward implementation of the grid into a 3D code already
employing spherical polar coordinates.  The Yin-Yang grid is
successfully used on massively parallel supercomputers in the field of
geophysical science for simulations of mantle convection and the
geodynamo. In these applications the thermal convection equation and
the magnetohydrodynamic (MHD) equations are solved on the Yin-Yang
grid using a second-order accurate finite difference method. Here, we
also adopt the Yin-Yang grid, and use it for astrophysically relevant
(finite-volume) hydrodynamic simulations for the first time.

The paper is structured as follows. In section 2, we describe the
basics of the Yin-Yang grid configuration including the
transformations of coordinates and vectors between the Yin and Yang
grid patches. In section 3, we provide the details of the
implementation of the Yin-Yang grid into the PROMETHEUS hydrodynamic
code, and also discuss the resulting necessary modifications of its 3D
gravity solver that is based on spherical harmonics.  In section 4, we
present the results of the test calculations we have performed
including a test with self-gravity. In section 5, we discuss the
conservation problem arising when applying the Yin-Yang grid. Then we
report on the efficiency and performance gain obtained with the
Yin-Yang grid compared to a spherical polar grid in section
6. Finally, we give the conclusions from our study in section 7.

\section{Yin-Yang Grid}
\begin{figure}
\includegraphics[width=0.5\textwidth]
{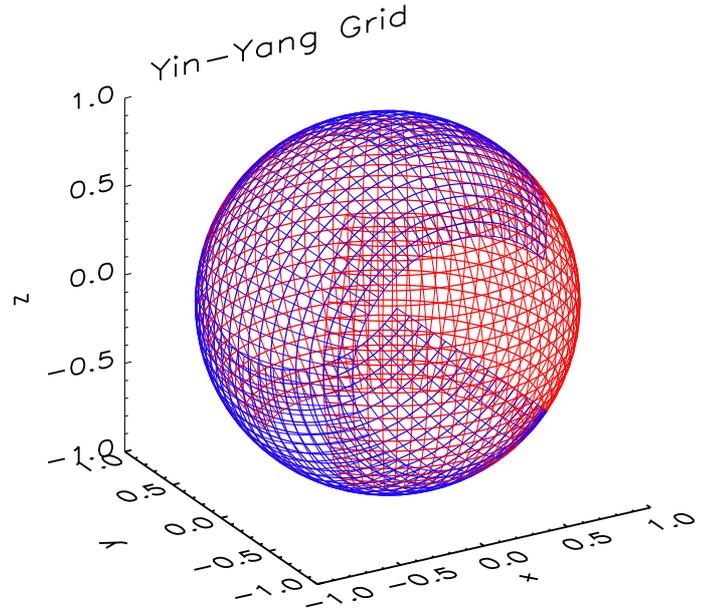}
\caption{An axis-free Yin-Yang grid configuration plotted on a
  spherical surface. Both the Yin (red) and Yang (blue) grid are the
  low latitude part of the normal spherical polar grid and are
  identical in geometry. The Yang grid is obtained from the Yin grid
  by two rotations, and vice versa.}
\label{fig:YY1}
\end{figure}

The Yin-Yang grid configuration is shown in Fig. \ref{fig:YY1}. Both
the Yin and the Yang grid are simply a part of a usual spherical polar
grid and are identical in geometry. The angular domain of each grid
patch is given by
\begin{equation}
\theta=\left[ \frac{ \pi}{4}-\delta, \,
              \frac{3\pi}{4}+\delta\right] \;\cap\; 
\phi  =\left[-\frac{3\pi}{4}-\delta,\,
              \frac{3\pi}{4}+\delta\right]
\label{eq:YYdom}
\end{equation}
where $\theta$ and $\phi$ are the colatitude and azimuth,
respectively.  Note that it is necessary to add at least one extra
buffer grid zone to both angular directions in order to ensure an
appropriate overlap of the grids.  The angular width $\delta$ of this
buffer zone depends on the grid resolution, \ie $\delta \equiv \Delta
\theta = \Delta \phi$, where for simplicity we assumed equal angular
spacing in $\theta$- and $\phi$-direction.  The angular domain is
hereby extended by $2\delta$ in both angular directions.  The Yin and
Yang grid are patched together in a specific manner forming a
spherical shell with a small overlapping region covering approximately
$6\%$ of a sphere's surface. Stacking up Yin-Yang shells in radial
direction results in a 3D grid that is identical to the usual
spherical polar grid in radial direction. It is obvious that, unlike
in the case of the spherical polar grid, the problematic high latitude
sections of the sphere are avoided, and the angular zoning is almost
equidistant.

The Cartesian coordinates 
\begin{equation}\label{eq:cc-yin}
  (x^{(n)}, y^{(n)}, z^{(n)}) = (r \sin \theta^{(n)} \cos \phi^{(n)},\,
                              r \sin \theta^{(n)} \sin \phi^{(n)},\,
                              r \cos \theta^{(n)} )
\end{equation}
corresponding to the Yin grid, denoted by a superscript $(n)$, and the
Cartesian coordinates 
\begin{equation}\label{eq:cc-yang}
  (x^{(e)}, y^{(e)}, z^{(e)}) = (r \sin \theta^{(e)} \cos \phi^{(e)},\,
                              r \sin \theta^{(e)} \sin \phi^{(e)},\,
                              r \cos \theta^{(e)} )
\end{equation}
corresponding to the Yang grid, denoted by a superscript $(e)$, are
related to each other through the transformation
\begin{equation}\label{eq:yy-tran1}
 \left(\begin{array}{c}
         x^{(e)} \\ y^{(e)} \\ z^{(e)}
       \end{array}\right) 
   = M
 \left(\begin{array}{c}
         x^{(n)} \\ y^{(n)} \\ z^{(n)}
      \end{array}\right) 
\end{equation}
where
\begin{equation}\label{eq:yy-tran2}
M=\left(\begin{array}{ccc}
            -1 & 0 & 0 \\
             0 & 0 & 1 \\
             0 & 1 & 0 
         \end{array}\right).
\end{equation}
This Yin-Yang coordinate transformation can also be considered as two
subsequent rotations. Accordingly, the transformation matrix $M$ can
be written as $R_x(90^\circ)\, R_z(180^\circ)$, where $R_x$ and $R_z$
are the transformation matrices of rotations by $90^\circ$ around the
$x$-axis and by $180^\circ$ around the $z$-axis in counterclockwise
direction, respectively. For the inverse transformation matrix $M^{-1}
= M$ holds.

The relation between the spherical coordinates of the Yin and Yang
grid patches can be derived directly from the transformation matrix
$M$. Because the Yin-Yang coordinate transformation involves only
rotations, it implies that the radial coordinate is identical on the
Yin and the Yang grid. The angular coordinates transform as
\begin{eqnarray}
  \theta^{(e)}&=&\arccos \left( \sin\theta^{(n)}\sin\phi^{(n)} \right),
\label{eq:yy-tran3} \\
  \phi^{(e)}&=&\arctan\left(\frac{\cos\theta^{(n)}}{-\sin\theta^{(n)}
                              \cos\phi^{(n)}}\right). 
\label{eq:yy-tran4}
\end{eqnarray} 
Note that the inverse transformation has the same form as
(\ref{eq:yy-tran3}) and (\ref{eq:yy-tran4}) but exchanging the (grid)
superscripts.
 
Vector components in spherical coordinates transform according to
\begin{equation}\label{eq:yy-tran5}
  \left(\begin{array}{c}
          v_r^{(e)} \\ v_\theta^{(e)} \\ v_\phi^{(e)}
        \end{array}\right) 
= P
  \left(\begin{array}{c}
          v_r^{(n)} \\ v_\theta^{(n)} \\ v_\phi^{(n)}
\end{array}\right) \,
\end{equation}
where
\begin{equation}\label{eq:yy-tran6}
P=\left(\begin{array}{ccc}
  1\, & 0 & 0 \\
  0\, & -\sin\phi^{(e)}   \sin\phi^{(n)} 
      & -\cos\phi^{(n)} / \sin\theta^{(e)}\\
  0\, &  \cos\phi^{(n)} / \sin\theta^{(e)}
      & -\sin\phi^{(e)}   \sin\phi^{(n)}
        \end{array}\right) 
\end{equation} 
is the vector transformation matrix. When switching (grid)
superscripts $(e)$ and $(n)$ in matrix $P$, the inverse vector
transformation matrix is obtained. For a detailed derivation of the
transformation matrix $P$, we refer to section\,3 of
\citet{basicyy}. Note that the vector transformation matrix $P$ is
singular at $\sin\theta^{(e)}=0$, but this singular point is
rectifiable. In practice, one can always decompose vectors into their
Cartesian components and perform the corresponding transformation.

\section{Implementation}
We have implemented the Yin-Yang grid into our explicit finite-volume
Eulerian hydrodynamics code, PROMETHEUS, which integrates the equations
of multidimensional hydrodynamics using the piecewise parabolic method
\citep[PPM;][]{PPM} and dimensional splitting. The code also includes
a Poisson solver based on spherical harmonics to handle self-gravity.

\subsection{Hydrodynamics solver}
Firstly, the Yin-Yang grid needs to be constructed. Since both the Yin
and the Yang grid are part of a spherical polar grid an analogous
spatial discretization in angular direction can be used. For example,
the $\theta$ and $\phi$ coordinates of the zone center of an angular
zone $(j, k)$ of a Yin-Yang grid, having $N_\theta$ zones in
$\theta$-direction and $N_\phi$ zones in $\phi$-direction, are given
by
\begin{eqnarray}
  \theta_j &=& \theta_{min}+j\Delta\theta-\frac{\Delta\theta}{2}
               \qquad {\rm for} \quad  1 \leq j \leq N_{\theta}, \\
  \phi_k   &=& \phi_{min}+k\Delta\phi-\frac{\Delta\phi}{2} 
               \qquad {\rm for} \quad  1 \leq k \leq N_{\phi}, 
\end{eqnarray}
where
\begin{eqnarray}
  \Delta\theta &=& \frac{\theta_{max}-\theta_{min}}{N_\theta},\\
  \Delta\phi   &=& \frac{\phi_{max}-\phi_{min}}{N_\phi}
\end{eqnarray}
are the respective angular grid spacings.

The range of values for the colatitude $\theta$ and the azimuth angle
$\phi$ are as given in (\ref{eq:YYdom}), and for simplicity we set
$\Delta \theta = \Delta \phi$.  In radial direction no modification is
required. The geometric property of the Yin-Yang grid allows us to
make use of the coordinate arrays $r_i$, $\theta_j,$ and $\phi_k$
twice by enforcing the same grid resolution for both grid
patches. This approach avoids doubling the coordinate arrays.

Only simple modifications are needed concerning the data and program
structure. Arrays with three spatial indices, e.g., $i$, $j$, and $k$,
need an extra grid index, say, $l$. For example, the array for the
density field will be $\rho(i,j,k,l)$ instead of $\rho(i,j,k)$. As a
consequence any triple loop running over indices $i$, $j$, and $k$ in
the program becomes a fourfold loop over $i$, $j$, $k$, and $l$
instead. Otherwise, the Yin-Yang grid allows one to exploit without
any further modification any already implemented finite-volume scheme
in spherical coordinates to solve the equations of hydrodynamics.

Different from the spherical polar grid, the Yin-Yang grid requires no
boundary conditions in angular directions. Each grid patch
communicates with its neighboring patch using information from ghost
zones that is obtained by interpolation of data between internal grid
zones of the neighboring grid patch. Interpolation is only required in
the two angular coordinates as the radial part of the Yin-Yang grid is
identical to that of a spherical polar grid. It is straightforward to
determine the corresponding interpolation coefficients.  The mapping
of vector quantities between the Yin and Yang grid patches requires an
additional step. After interpolating the vector components they must
be transformed according to the transformation given in
Eq.\,(\ref{eq:yy-tran5}) from the Yin to the Yang angular coordinate
system, and vice versa.

We tested two interpolation procedures. In the first one all primitive
state variables (density, velocity, energy, pressure, temperature,
abundances) are interpolated ignoring the resulting small
thermodynamic inconsistencies. In the second procedure, we only
interpolate the conserved quantities (density, momentum, total energy,
and abundances), and compute the velocity and the remaining
thermodynamic state variables consistently via the equation of
state. Both procedures produce very similar results which differ at
the level of the discretization errors. As the second procedure is
more consistent we use it as the standard one in our code.  

\begin{figure}
\includegraphics[width=0.5\textwidth]
{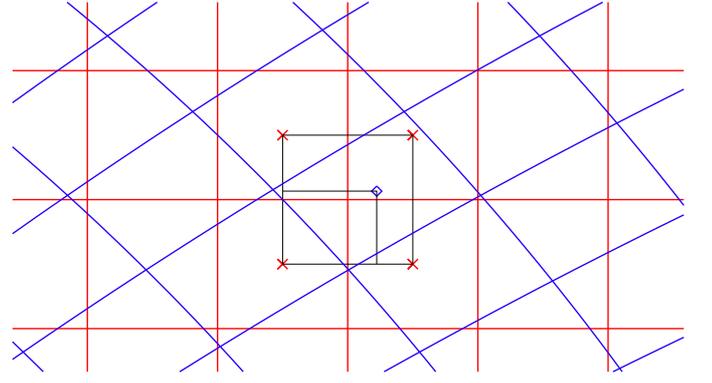}
\caption{A Mercator projection of an overlap region of the Yin-Yang
  grid. In case of bi-linear interpolation, four neighboring values of
  the underlying grid (red) will be used to determine the
  zone-centered value of a ghost zone in the grid on top (blue). The
  interpolation coefficients are determined by the relative distances,
  denoted by black lines, between the interpolation point (diamond)
  and the four neighboring points (crosses).}
\label{fig:YY2}
\end{figure}

An example of overlapping situations which are encountered when using
a Yin-Yang grid is shown in Fig.\,\ref{fig:YY2}. For simplicity, we
use bi-linear interpolation in order to prevent unwanted oscillation.
Because the grid patches are fixed in both angular directions the
interpolation coefficients for each ghost zone need to be calculated
only once per simulation at the initialization step. After
initialization, the coefficient map is stored in an array for later
usage. Moreover, the symmetry property of the Yin-Yang transformation
allows one to make use of the interpolation coefficients twice for
both grids. 

Because the Yin-Yang grid is an overlapping grid integral quantities
such as the total mass or total energy on the computational domain
cannot be obtained by just summing local quantities from every grid
cells. Doing so will result in counting the contributions in the
overlapping region twice. To circumvent this problem, weights are
given to each grid zone during the summation. Suppose a grid zone has
an overlapping volume fraction $\alpha$ the cell will receive a weight
$w=1.0-0.5\alpha$. Zones in the non-overlapping region receive the
weight of $1.0$, i.e., the entire zone contributes to the integral
while, on the other hand, zones that are fully contained within the
overlapping region have a weight $0.5$.  The volume fraction $\alpha$
does not depend on the radial coordinate and can be thought of as an
area fraction since the grid patches are not offset in radial
direction.  Prior to the area integration, one needs to determine for
each zone interface of the underlying grid the points where the
interface is intersected by the boundary lines of the other grid,
e.g., points on the Yin grid intersected by the boundary lines of the
Yang grid. The intersection points can be determined using the
Yin-Yang coordinate transformation in (\ref{eq:yy-tran3}) and
(\ref{eq:yy-tran4}), respectively. The integration in the overlapping
area is then carried out using the trapezoidal method. This procedure
is also described in \citet{conserv}. Once the area or volume fraction
$\alpha$ is calculated, the weights for each cell are obtained
easily. Note that these weights need to be calculated only at the
initialization step, and are stored for later usage in a coefficient
map $w(j,k)$, where $j$ and $k$ are the indices referring to the
$\theta$ and $\phi$ coordinates, respectively. The coefficient map can
be applied to both grids without any modification. Using the above
described approach, the volume or surface area of the grids can be
calculated with an accuracy up to machine precision.

\subsection{Gravity solver}
\label{sec.gravpot}
The 3D Newtonian gravitational potential is computed from Poisson's
equation in its integral form using an expansion into spherical
harmonics as described in \citet{grav-solver}. Because the algorithm
of these authors is based on a (single) spherical polar grid the
density on the Yin-Yang sphere has to be interpolated onto an
auxiliary spherical polar grid. The interpolation used is first-order
accurate, and due to the simplicity of the Yin-Yang grid configuration
has to be performed only in the two angular dimensions. Concerning the
resolution of the auxiliary grid, it is natural to employ the same
grid resolution as that used for the Yin-Yang grid in all three
spatial dimensions.  The orientation of the auxiliary grid can be
chosen freely in principle. However, it is convenient to align it with
one of the two grid patches (the Yin-grid in our case). Once the
density is interpolated onto the auxiliary spherical grid we compute
the gravitational potential, as suggested by \citet{grav-solver}, at
zone interfaces instead of at zone centers on both the Yin and Yang
grid. The gravitational acceleration at zone centers can then be
obtained by central differencing the potential. Note that the
interpolation coefficients for the density need to be calculated only
once per simulation, because both the auxiliary grid and the Yin-Yang
grid are fixed in angular directions.  In addition, all angular
weights, Legendre polynomials, and their integrals required for the
calculation of the gravitational potential are stored after the
initialization step for later usage.

It is also possible to directly calculate the gravitational
acceleration at zone centers. The gravitational potential is given by
(see Eqs.\ (5), (6) and (7) in \citet{grav-solver})
\begin{equation}
 \Phi(r,\theta,\phi) = -G\sum_{l=0}^{\infty} \frac{4\pi}{2l+1}
                        \sum_{m=-l}^{l} Y^{lm} (\theta,\phi)
                        \left(\frac{1}{r^{l+1}}C^{lm}(r) +
                              r^lD^{lm}(r) \right)
\label{eq:pot}
\end{equation}
with
\begin{eqnarray}
  C^{lm}(r) = \int\limits_{4\pi}  d\, \Omega^\prime 
                                Y^{lm*}(\theta^\prime, \phi^\prime)
             \int\limits_{0}^{r} dr^\prime r^{\prime l+2} 
             \rho(r^\prime, \theta^\prime, \phi^\prime) \, , \\
  D^{lm}(r) = \int\limits_{4\pi} d\, \Omega^\prime 
                               Y^{lm*}(\theta^\prime, \phi^\prime)
             \int\limits_{r}^{\infty} dr^\prime r^{\prime 1-l}
             \rho(r^\prime, \theta^\prime, \phi^\prime) \, ,
\end{eqnarray}
where $Y^{lm}$ and $Y^{lm*}$ are the spherical harmonics and their
complex conjugates, $\rho$ is the density, and $d\,\Omega\equiv
\sin{\theta}\,d\theta\,d\phi$. The gravitational acceleration in
radial direction is then
\begin{eqnarray}
  &&\frac{\partial}{\partial r}\Phi(r,\theta,\phi)
  = \nonumber \\
  &&-G \sum_{l=0}^{\infty} \frac{4\pi}{2l+1} 
     \sum_{m=-l}^{l} Y^{lm} (\theta,\phi) \frac{d}{dr}
     \left(\frac{1}{r^{l+1}} C^{lm}(r) +
           r^l D^{lm}(r) \right) \, .
\label{eq:accel_r-1}
\end{eqnarray}
Writing the radial derivative in Eq.\,(\ref{eq:accel_r-1}) as
\begin{eqnarray}
  \frac{d}{dr} \left( \frac{1}{r^{l+1}}C^{lm}(r)+r^lD^{lm}(r) \right) &=&
  \frac{1}{r^{l+1}} \frac{d}{dr}C^{lm}(r) -
  \frac{l+1}{r} \cdot \frac{1}{r^{l+1}} C^{lm}(r) \nonumber \\
  &+& r^l \frac{d}{dr} D^{lm}(r) + 
  \frac{l}{r} \cdot r^{l} D^{lm}(r) \, ,
\label{eq:accel_r-2}
\end{eqnarray}
and noticing that the first and third term on the right hand side of
this expression cancel each other because of the identities
\begin{equation}
 \frac{d}{dx} \int\limits_{0}^{x} f(x^\prime) dx^\prime = f(x)
\end{equation}
and
\begin{equation}
  \frac{d}{dx} \int\limits_{x}^{\infty} f(x^\prime) dx^\prime = -f(x)
\end{equation}
the gravitational acceleration in radial direction becomes
\begin{eqnarray}
  \frac{\partial}{\partial r} \Phi(r,\theta,\phi)
  = 
  &-&G \sum_{l=0}^{\infty} \frac{4\pi}{2l+1} 
  \sum_{m=-l}^{l} Y^{lm} (\theta,\phi) \nonumber \\
  &&\left( -\frac{l+1}{r} \cdot \frac{1}{r^{l+1}} C^{lm}(r) +
          \frac{l}{r} \cdot r^l D^{lm}(r) \right) \, .
\label{eq:accel_r}
\end{eqnarray}

The corresponding expressions for the gravitational acceleration in
the two angular directions are easy to obtain since the spherical
harmonics $Y^{lm}$ are the only angular-dependent terms in
Eq.\,(\ref{eq:pot}). Therefore, we only need to consider the partial
derivatives of the spherical harmonics with respect to the $\theta$
and $\phi$ coordinates. As the spherical harmonics are given by
\begin{equation}
  Y^{lm}(\theta,\phi) = N^{lm} P^{lm}(\cos{\theta}) e^{im\phi} \, ,
\end{equation}
where $N^{lm}$ is the normalization constant and $P^{lm}$ the 
associated Legendre polynomial, one finds
\begin{equation}
  \frac{\partial}{\partial\theta} Y^{lm}(\theta,\phi)
  = N^{lm} e^{im\phi} \frac{d}{d\theta} P^{lm}(\cos{\theta})
\end{equation}
and
\begin{equation}
  \frac{\partial}{\partial\phi} Y^{lm}(\theta,\phi)
  = i m Y^{lm}(\theta,\phi).
\end{equation}
The derivatives of the associated Legendre polynomials are easily
obtained using the recurrence formula
\begin{equation}
 (x^2-1)\frac{d}{dx} P_l^m(x) = l x P_l^m(x) - (l+m) P_{l-1}^m(x) \, .
\end{equation}
Thus, one finds for the gravitational acceleration in the angular
directions
\begin{eqnarray}
  \frac{1}{r} \frac{\partial}{\partial\theta} \Phi(r,\theta,\phi)
  =
  -\frac{G}{r} \sum_{l=0}^{\infty} \frac{4\pi}{2l+1}
  &&\sum_{m=-l}^{l} N^{lm} e^{im\phi} \frac{d}{d\theta}
  P^{lm}(\cos{\theta})
  \nonumber \\
  &&\left( \frac{1}{r^{l+1}} C^{lm}(r) + r^l D^{lm}(r) \right)
\label{eq:accel_theta}
\end{eqnarray}
and
\begin{eqnarray}
  \frac{1}{r\sin{\theta}} \frac{\partial}{\partial\phi} \Phi(r,\theta,\phi)
  =
  -\frac{G}{r\sin{\theta}} &&\sum_{l=0}^{\infty} \frac{4\pi}{2l+1}
  \sum_{m=-l}^{l} i m Y^{lm}(\theta,\phi) \nonumber \\
  &&\left( \frac{1}{r^{l+1}} C^{lm}(r) + r^l D^{lm}(r) \right) \, .
\label{eq:accel_phi}
\end{eqnarray}
Obviously, the expressions for three components of the gravitational
acceleration (see Eqs.\,(\ref{eq:accel_r}), (\ref{eq:accel_theta}),
and (\ref{eq:accel_phi})) are similar to that for the gravitational
potential itself (see Eq.\,(\ref{eq:pot})).  Hence, besides computing
derivatives of Legendre polynomials, our extended Poisson solver can
provide without much additional effort both the gravitational
potential and the corresponding acceleration.

Usage of the analytic expressions for the gravitational acceleration
avoids the errors arising from the numerical differentiation of the
gravitational potential.  However, tests show that the results
obtained using either the gravitational potential computed with the
``standard'' Poisson solver and subsequent numerical differentiation
or directly the gravitational acceleration provided by the extended
Poisson solver differ only very slightly (see next section). Thus, we
decided to stick to the ``standard'' Poisson solver in our simulations
and compute the gravitational acceleration by numerical
differentiation, as it requires no modification of our code.

\begin{figure}
\includegraphics[width=0.5\textwidth]{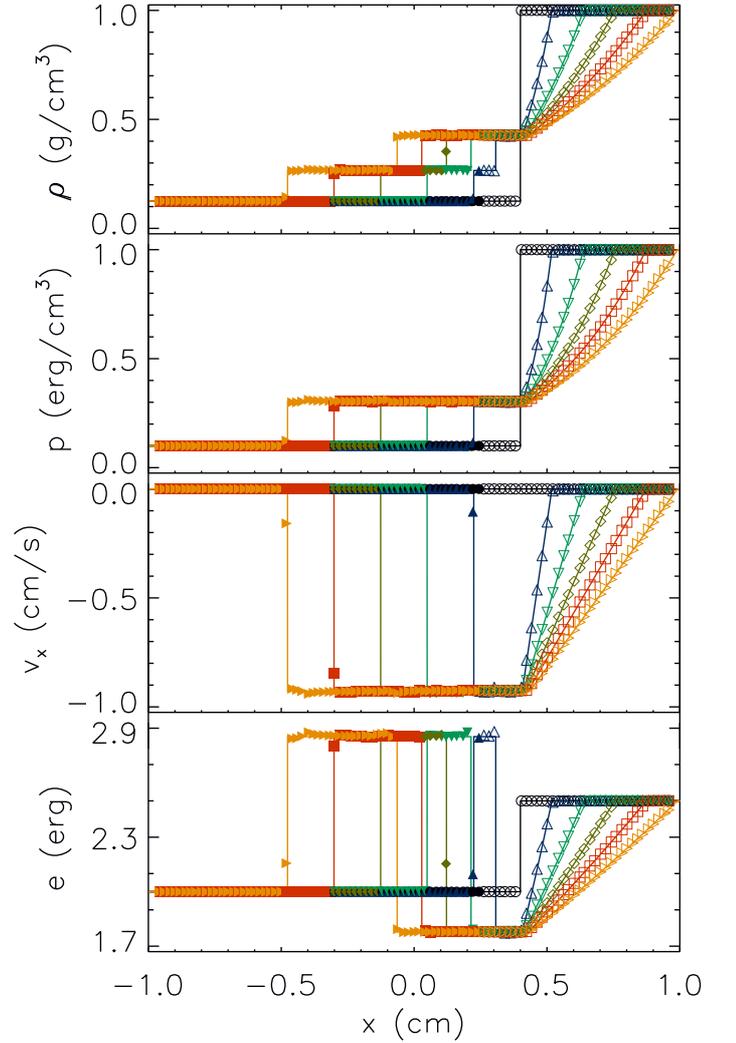}
\caption{One dimensional profiles of density $\rho$, pressure $p$,
  velocity in $x$-direction $v_x$, and specific internal energy $e$
  are shown along the $x$-direction at $z^{(n)}=0.25\,$cm and
  $y^{(n)}=0\,$cm (dashed-dotted line in Fig.\,\ref{fig:SODSPH-2}) for
  the shock tube simulation at every $0.1\,$s. Open and filled symbols
  represent data points on the Yin and Yang grid, respectively. Solid
  lines give the distributions calculated with an exact Riemann
  solver.}
\label{fig:SODSPH-1}
\end{figure}

\section{Test Suites}

\subsection{Sod Shock Tube}
The first problem of our test suite is the planar Sod shock tube
problem, a classical hydrodynamic test problem \citep{sod}. We
simulated this (1D Cartesian) flow problem using spherical coordinates
and the Yin-Yang grid.  The initial state consists of two constant
states given by
\begin{equation}\label{sod1}
(\rho,p,v_x) = \left\{\begin{array}{ll}
                (1.0,   1.0, 0.0) & \mbox{if $x^{(n)}>    0.4\,$cm} \\ 
                (0.125, 0.1, 0.0) & \mbox{if $x^{(n)}\leq 0.4\,$cm}
               \end{array}\right.,
\end{equation} 
where $\rho$, $p$, and $v_x$ are the density, pressure and the
velocity in $x$-direction of the fluid, respectively. We assume the
fluid to obey an ideal gas equation of state with an adiabatic index
$\gamma=1.4$.  The surface separating the two constant states is a
plane orthogonal to the $x$-axis located at $x^{(n)}=0.4\,$cm, the
(positive) $x$-axis corresponding to a radial ray with angular
coordinates $\theta^{(n)}=\pi/2$ and $\phi^{(n)}=0$. Thus, this 1D
planar Sod shock tube problem invokes all three spherical velocity
components $v_r$, $v_{\theta}$, and $v_{\phi}$ when simulating the
flow in spherical polar coordinates. This allows us to test both the
scalar and vector transformations as well as the interpolation between
the Yin and Yang grid patches. The simulation was carried out on an
equidistant Yin-Yang grid of $400\, (r) \times 92\, (\theta) \times
272\, (\phi) \times 2$ zones (i.e., with an angular resolution of one
degree; see Eq.(\ref{eq:YYdom})). In radial direction the
computational domain ranges from $r=0.05\,$cm to $r=1.0\,$cm.  We
impose a zero-gradient boundary condition at both edges of the radial
domain.

\begin{figure}
\includegraphics[width=0.5\textwidth]{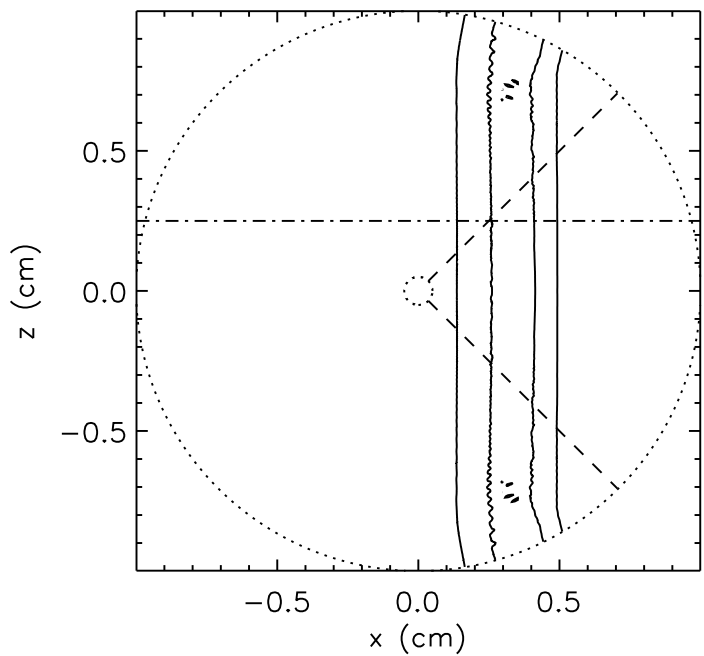}
\caption{Snapshot of density contours in the meridional plane
  $\phi^{(n)}=0$ at $t=0.15\,$s for the shock tube test
  problem. Dashed lines mark the Yin-Yang grid boundary, while the
  dotted circular curves represent the inner and outer radial boundary
  of the computational domain, respectively.  The one dimensional
  profiles shown in Fig.\,\ref{fig:SODSPH-1} are re-sampled along the
  dashed-dotted line at $z^{(n)}=0.25\,$cm.}
\label{fig:SODSPH-2}
\end{figure}

The solution of the shock tube problem is well-known. We compare our
results with the solution calculated using an exact Riemann solver
\citep{toro}. For comparison, data are re-sampled along the
$x$-direction with a spacing $\Delta x = 0.002\,$cm.
Fig.\,\ref{fig:SODSPH-1} shows one dimensional profiles of $\rho$,
$p$, $v_x$, and $e$ (specific internal energy), respectively, along
the $x$-direction at $z^{(n)}=0.25\,$cm and $y^{(n)}=0\,$cm
(dashed-dotted line in Fig.\,\ref{fig:SODSPH-2}) at different
times. Our results agree very well with the solution obtained with the
exact Riemann solver. The grid resolution is sufficiently high to give
a sharp shock front and contact discontinuity while the rarefaction
wave is smooth. The shock position is correct at all time throughout
the simulation. The re-sampled data yield an accuracy of approximately
$6\%$ on average for shock positions. The shock wave and the contact
discontinuity propagate smoothly across the Yin-Yang boundary located
at $x^{(n)}=0.25\,$cm without any noticeable effect by the existence
of the boundary. To illustrate this behavior, Fig.\,\ref{fig:SODSPH-2}
shows lines of constant density in the meridional plane $\phi^{(n)}=0$
at time $t=0.15\,$s. The isocontours are nearly perfectly straight
lines perpendicular to the $x$-axis that are unaffected by the
Yin-Yang boundary (dashed line). The contour lines are slightly bent
near the outer radial edge of the computational domain due to the
zero-gradient boundary condition we have imposed there.

\begin{figure*}
\includegraphics[width=0.5\textwidth]{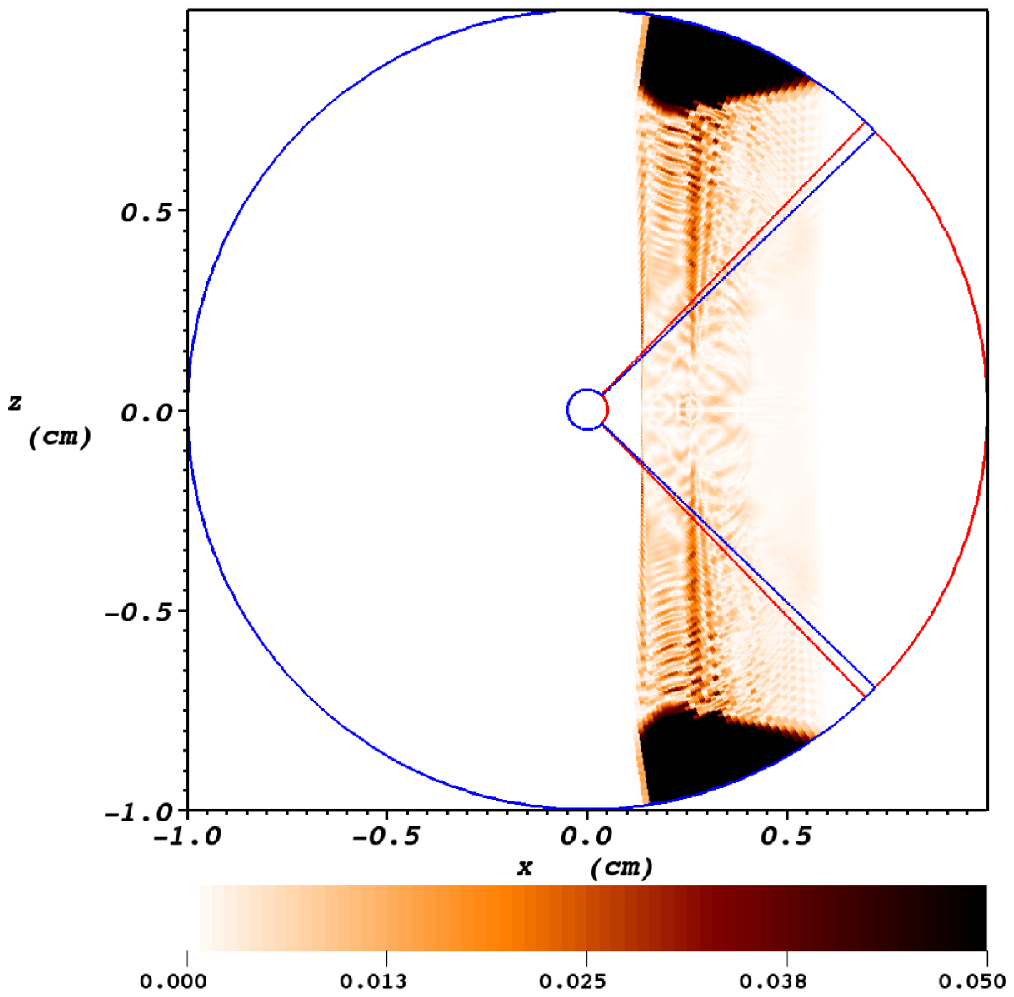}
\includegraphics[width=0.5\textwidth]{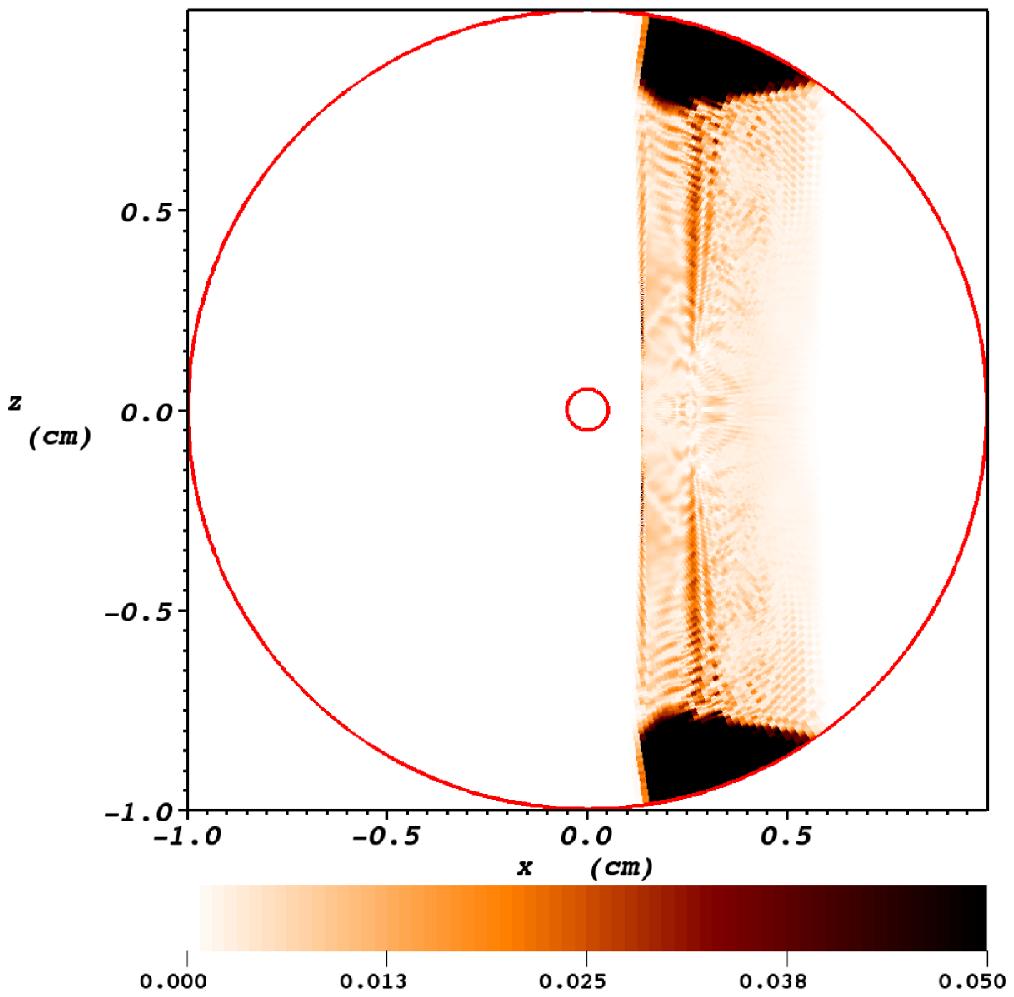}
\caption{Color maps of the tangential velocity defined by $\sqrt{
    (v_y^{(n)})^2 + (v_z^{(n)})^2 }$ in the meridional plane
  $\phi^{(n)}=0$ resulting from the (1D Cartesian) shock tube
  problem. The snapshots are computed using the Yin-Yang grid (left)
  and a standard spherical polar grid (right) at a time
  $t=0.15\,$s. On the left panel, red and blue lines mark the
  boundaries of the Yin and the Yang grid patches, respectively. On
  the right panel, the two red circles show the inner and outer
  boundary in the radial direction of the standard spherical polar
  grid. The labels at the color bars give the tangential velocity in
  units of cm/s. The color range is limited to $0.05\,$cm/s to
  emphasize the smallness of the tangential velocity far from the
  outer radial grid boundary.}
\label{fig:SODSPHVt}
\end{figure*}

In order to firmly demonstrate that the Yin-Yang boundary does not
cause numerical artifacts, we also computed this shock tube problem
with a standard spherical polar grid using the same radial and angular
resolution as for the Yin-Yang grid described above, \ie $400\, (r)
\times\, 180 (\theta) \times 360\, (\phi)$.  We imposed reflecting
boundary conditions in $\theta$-direction and periodic ones in
$\phi$-direction.  Fig.\,\ref{fig:SODSPHVt} shows a comparison of the
results obtained with both simulations. The two panels give the
tangential velocity, defined as $\sqrt{(v_y^{(n)})^2 +
  (v_z^{(n)})^2}$, in the meridional plane $\phi^{(n)}=0$ at time
$t=0.15\,$s for the Yin-Yang grid (left), and the standard spherical
polar grid (right), respectively. This velocity component should
remain exactly zero because of the chosen initial conditions. Thus, it
is a sensitive indicator whether the Yin-Yang boundary works properly,
which obviously is indeed the case as the left panel of
Fig.\,\ref{fig:SODSPHVt} shows no hint of the location of that
boundary.  The modulus of the tangential velocity does nowhere exceed
a value of $0.05\,$cm/s or approximately $5\%$ of the shock velocity
(in $x$-direction) except near the outer radial edge of the grids,
where the boundary condition causes larger numerical errors.  Note
that nonzero tangential velocities are encountered on both the
Yin-Yang grid and the standard spherical polar grid in the same grid
regions at the same level. We thus conclude that they are the result
of numerical errors that unavoidably occur when propagating a planar
shock across a spherical polar grid, be it a standard one or a
Yin-Yang grid.

\subsection{Taylor-Sedov Explosion}
As a second test for our code we consider the Taylor-Sedov explosion
problem. We set up the initial state for the problem by mapping a
spherically symmetric analytic solution \citep{landau} onto the
computational grid. We choose the parameters of the problem to mimic a
supernova explosion in an interstellar medium. Because the shock wave
resulting from the explosion is spherically symmetric with respect to
the center of the explosion, we assume the explosion center to be
located at the point $(x^{(n)}, y^{(n)}, z^{(n)}) = (7.0, 0.0, 2.5)
\times 10^{19}\,$cm.  Hence, this second test problem also involves a
non-zero flux of mass, momentum, and energy across the Yin-Yang
boundary, and as the previous shock tube test, it probes whether that
boundary causes any numerical artifacts.

The initial shock radius is $r_0 = 2.9625 \times 10^{19}\,$cm
orresponding to a time $t_{exp} = 0.34\times 10^{11}\,$s past the
onset of the explosion, and the explosion energy was set to $E_0 =
10^{51}\,$erg. The ambient medium into which the shock wave is
propagating is at rest. It has a constant density $\rho_b =
10^{-25}\,$g/cm$^3$, and a constant pressure $p_b = 1.4\times
10^{-13}\,$erg/cm$^3$. The fluid is described by an ideal gas equation
of state with an adiabatic index $\gamma = 5/3$, resulting in a
density jump across the shock front of $(\gamma+1) / (\gamma-1) =
4$. We use a grid resolution of $400 \times 92 \times 272 \times 2$
zones, a computational domain covering the radial interval $r = [0.5,
  15.] \times 10^{19}\,$cm, and employ a zero-gradient boundary
condition at both the inner and the outer radial boundary.

\begin{figure}
\includegraphics[width=0.5\textwidth]{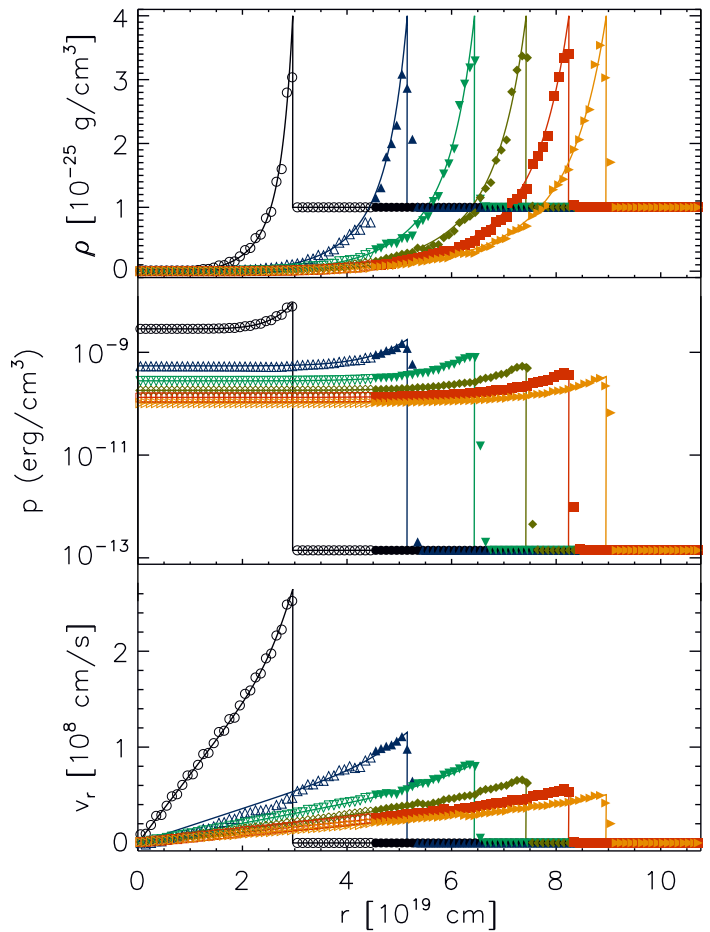}
\caption{Distributions of density (top), pressure (middle) and radial
  velocity (bottom) versus radius from the explosion center (located
  at $(x^{(n)},y^{(n)},z^{(n)})= (7.0, 0.0, 2.5) \times 10^{19}\,$cm
  for the Taylor-Sedov explosion problem plotted at every
  $10^{11}\,$s. Open symbols are data points from the Yin grid, while
  filled symbols represent sampled data from the Yang grid. The solid
  lines give the corresponding analytic solution.  The data are
  re-sampled along the dashed-dotted line shown in
  Fig.\,\ref{fig:SEDOVOFF-2}.}
\label{fig:SEDOVOFF-1}
\end{figure}

\begin{figure}
\includegraphics[width=0.5\textwidth]{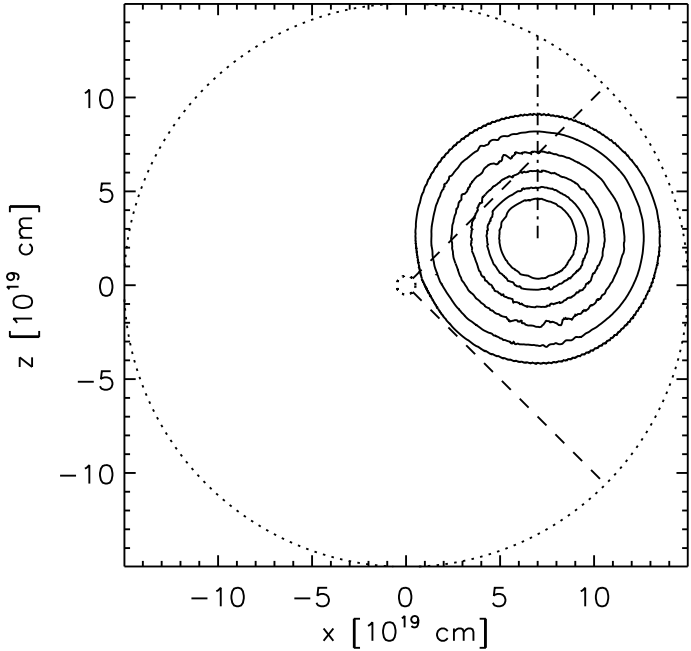}
\caption{Lines of constant density in the meridional plane $\phi^{(n)}
  = 0$ obtained from our simulation of a Taylor-Sedov explosion. The
  snapshot is taken at a simulation time $t_{sim}=2.0 \times
  10^{11}\,$s which corresponds to an explosion time $t_{exp}\approx
  2.34\times 10^{11}\,$s. The dashed lines mark the Yin-Yang boundary,
  while the two dotted circles represent the inner and outer radial
  boundary of the computational domain, respectively.  The data
  presented in Fig.\,\ref{fig:SEDOVOFF-1} are re-sampled along the
  dashed-dotted line .}
\label{fig:SEDOVOFF-2}
\end{figure}

Our results are shown together with the analytic solution in
Fig.\,\ref{fig:SEDOVOFF-1}. We have re-sampled our data and calculated
radial profiles of the density $\rho$, pressure $p$, and radial
velocity $v_r$ along a line in $z$-direction through the explosion
center using a uniform radial spacing $\Delta r = 10^{18}\,$cm. As one
can see the numerical results agree very well with the analytic
solution. All flow quantities are smooth across the Yin-Yang boundary,
\ie the shock wave passes that boundary without any noticeable
numerical artifact. Due to the finite resolution the density jump
across the shock front is slightly smaller in the simulation than the
analytic value of four. However, the shock front is sharp throughout
the whole simulation, and it propagates with the correct speed. One
distinct feature of the Taylor-Sedov solution is its spherical
symmetry. To illustrate that the Yin-Yang grid does not destroy this
symmetry of the solution, we show a set of lines of constant density
in the meridional plane $\phi^{(n)}=0$ in Fig.\,\ref{fig:SEDOVOFF-2}.
We also marked the line (dashed-dotted) along which the data given in
Fig.\, \ref{fig:SEDOVOFF-1} are re-sampled. The contour lines, all of
which are almost perfectly circular, are drawn at a simulation time
$t_{sim} = 2.0 \times 10^{11}\,$s (\ie time step number 1276)
corresponding to an explosion time $t_{exp} \approx 2.34\times
10^{11}\,$s.

\begin{figure}
\includegraphics[width=0.5\textwidth]{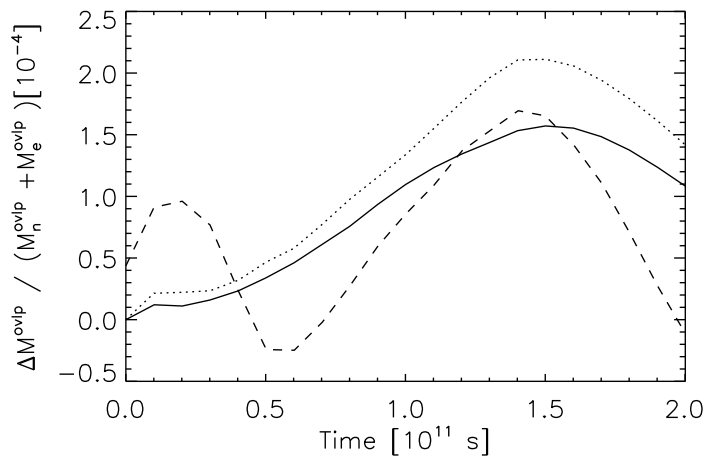}
\caption{Evolution of the mass within the overlap region for the
  Taylor-Sedov test case computed on the Yin grid, $M^{ovlp}_{n}$
  minus the mass computed on the Yang grid, $M^{ovlp}_{e}$, divided by
  the sum of these two masses. The dashed, dotted and solid lines give
  the solutions computed on a grid of $400 \times 32 \times 92 \times
  2$ zones, (\ie $3^\circ$ angular resolution), $400 \times 92 \times
  272 \times 2$ zones (\ie $1^\circ$ angular resolution), and $400
  \times 182 \times 542 \times 2$ zones (\ie $0.5^\circ$ angular
  resolution), respectively.}
\label{fig:SEDOVOFF-3}
\end{figure}

We further studied how the solution differs in the region where the
Yin and Yang grid overlap. To this end we compare the total mass
within the overlap region computed on the Yin and the Yang grid,
respectively. Fig.\,\ref{fig:SEDOVOFF-3} shows the evolution of the
relative mass difference, \ie the mass within the overlap region
computed on the Yin grid, $M^{ovlp}_{n}$ minus the mass computed on
the Yang grid, $M^{ovlp}_{e}$, divided by the sum of these two
masses. We calculated this quantity for three different (angular)
grids with $400 \times 32 \times 92 \times 2$ zones (\ie $3^\circ$
angular resolution), $400 \times 92 \times 272 \times 2$ zones (\ie
$1^\circ$ angular resolution), and $400 \times 182 \times 542 \times
2$ (\ie $0.5^\circ$ angular resolution), respectively.  For all three
grid resolutions the relative mass difference has a value of about
$10^{-4}$. Although its evolution with time is different in case of
the $3^\circ$ simulation (because the coarse angular grid causes large
errors when mapping the analytic initial data onto the grid which
determine the further evolution), Fig.\,\ref{fig:SEDOVOFF-3} shows
that for an angular resolution better than $1^\circ$ the relative mass
difference behaves similarly, its maximum value decreasing from
$2.1\times 10^{-4}$ at $1^\circ$ angular resolution to $1.5\times
10^{-4}$ at $0.5^\circ$ angular resolution. 

\begin{figure*}
\includegraphics[width=0.5\textwidth]{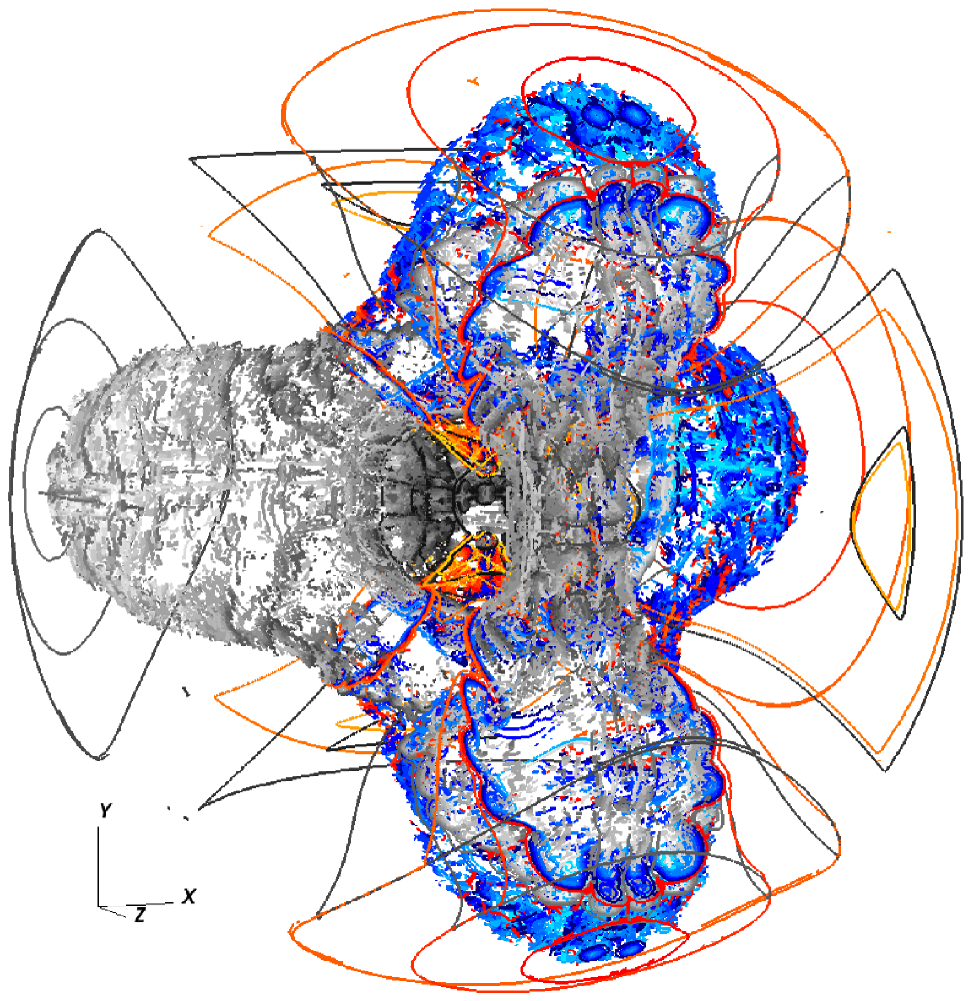}
\includegraphics[width=0.5\textwidth]{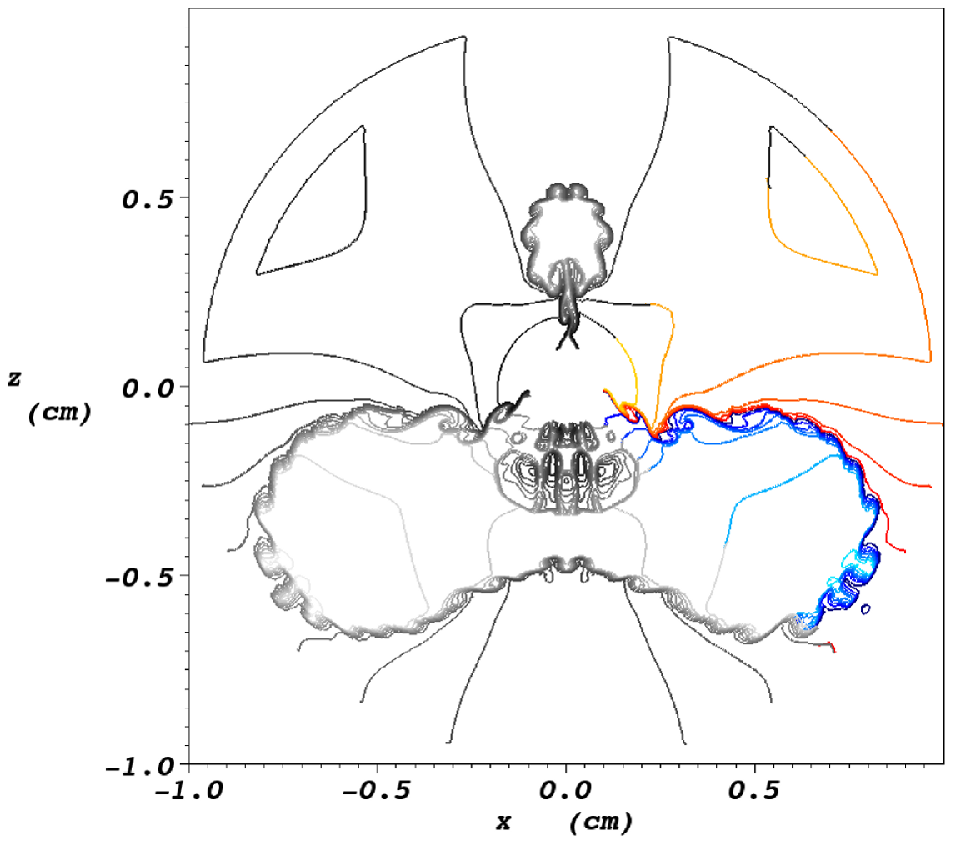}
\caption{Surfaces of constant density in 3D (left) and 2D (right;
  meridional cut at $\phi^{(n)}=0$) resulting from the simulation of
  the Rayleigh-Taylor instability described in the text at
  $t=2.85\,$s.  Contour lines on the Yin grid are shown using the
  blue-yellow colors while contour lines on the Yang grid are
  displayed using the white-black colors.}
\label{fig:RTCON}
\end{figure*}

\subsection{Rayleigh-Taylor Instability}
We also simulated a single mode Rayleigh-Taylor instability (RTI) on a
Yin-Yang sphere. The initial configuration consists of a spherical
shell of a heavier fluid of density $\rho_H = 2\,$g/cm$^3$ that is
supported against a constant gravitational field $g = 1\,$cm/s$^2$
pointing in negative radial direction by a spherical shell of a
lighter fluid of density $\rho_L = 1\,$g/cm$^3$. The boundary between
the two fluid shells is initially located at a radius $r=0.5\,$cm. To
balance the gravitational force, the initial (radial) pressure
distribution is set to
\begin{equation}\label{RT1}
P(r) = \left\{\begin{array}{ll}
                P_0      + g \rho_H\,(1.0-r) & \mbox{if\, $r\geq 0.5$ cm} \\ 
                P(r=0.5) + g \rho_L\,(0.5-r) & \mbox{if\, $r< 0.5$ cm}
               \end{array}\right.
\end{equation}
where $P_0=1\,$erg/cm$^3$. A radial velocity varying in angular
direction as the spherical harmonics $Y_l^m(\theta, \phi)$ with $l=3$
and $m=2$ is used to perturb the initial configuration. The amplitude
of the velocity perturbation is $2.5\%$ of the local sound speed
$c_s(r)$. Hence, the initial radial velocity is given by
\begin{equation}
  v_r(r,\theta,\phi) = -0.025\times c_s(r)\, Y_3^2(\theta,\phi).
\label{RT2}
\end{equation}
The spherical harmonics $Y_l^m(\theta,\phi)$ are connected with the
associated Legendre polynomials $P_l^m$ via the expression
\begin{equation}
  Y_l^m(\theta,\phi) = \sqrt{\frac{(l-m)!}{(l+m)!}} \, 
                       P_l^m(\cos\theta)e^{im\phi} \, .
\end{equation}
The perturbation mode $(l,m)=(3,2)$ yields a maximum radial velocity
in the directions 
\begin{equation}
  (\theta,\phi) = \left\{ (\pi-\alpha,  0    ), \,
                          (\pi-\alpha,  \pi  ), \,
                          (    \alpha,  \pi/2), \,
                          (    \alpha, -\pi/2)\right\} \, ,
\end{equation}
where $\alpha \equiv \arccos(\sqrt{3}/3)$. The remaining two velocity
components of the perturbation mode are set equal to $0$. The fluids
are described by an ideal gas equation of state with an adiabatic
index $\gamma = 1.4$. The simulation is carried out on a Yin-Yang grid
of $400 \times 92 \times 272 \times 2$ zones. To keep the fluid in
hydrostatic equilibrium, a zero-gradient boundary condition is used
for both the inner and outer boundary in radial direction. The inner
radial boundary is located at $r=0.1\,$cm.

\begin{figure}
\includegraphics[width=0.5\textwidth]{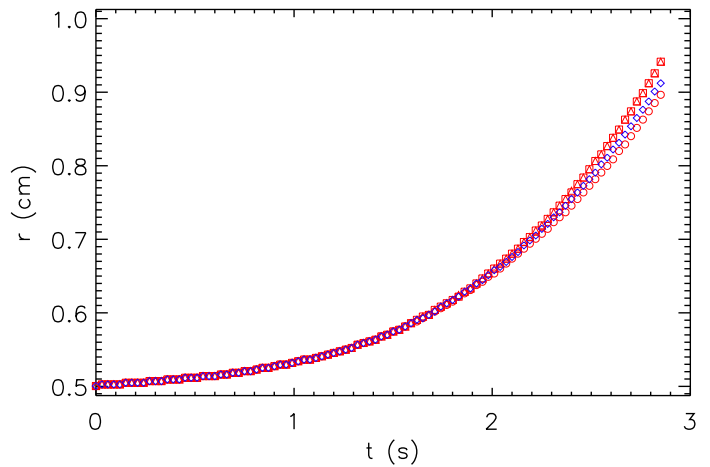}
\caption{Position of the heads of the RTI bubbles versus time.  Red
  symbols (circles, triangles, and squares) show data from the Yin
  grid, while blue symbols (diamonds) represent data on the Yang
  grid.}
\label{fig:RTGROWTH}
\end{figure}

\begin{figure*}
\includegraphics[width=\textwidth]{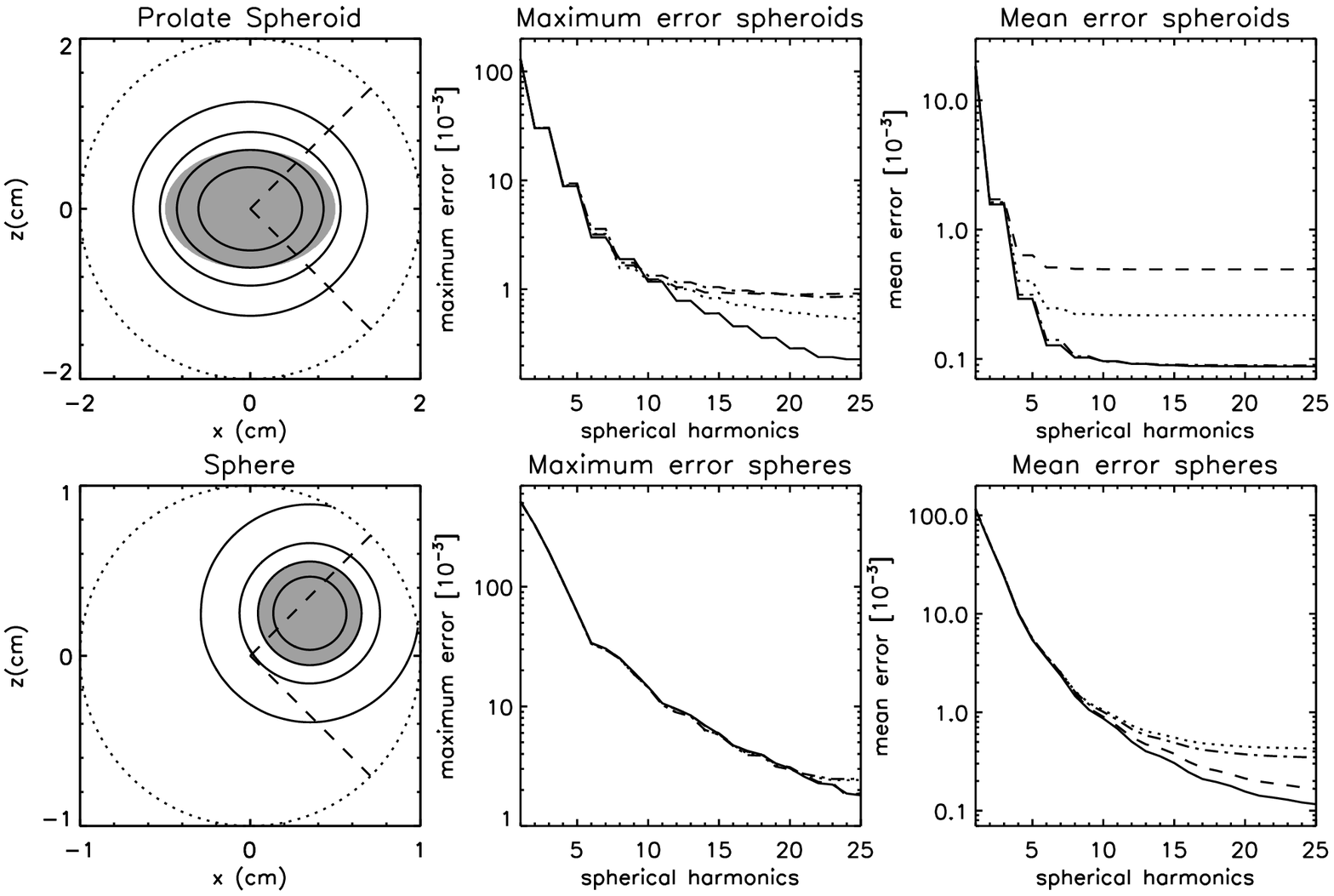}
\caption{Contour lines of the gravitational potential (left column)
  for two homogeneous self-gravitating configurations: a prolate
  spheroid (top row) with an axis ratio of $0.7$, and a sphere (bottom
  row). The configurations are indicated by the dark-gray shaded
  areas. Dashed lines show the Yin-Yang boundary, while dotted lines
  indicate the outer radial boundary of the computational grid.  The
  middle and right columns give the maximum and mean error of the
  numerically calculated gravitational potential for different grid
  resolutions as a function of the number of spherical harmonics used
  in our multipole gravity solver. The solid, dotted, dashed, and
  dashed-dotted lines in both columns correspond to a grid resolution
  of $400 \times 92 \times 272 \times 2$ zones , $400 \times 47 \times
  137 \times 2$ zones, $200 \times 92 \times 272 \times 2$ zones, and
  $200 \times 47 \times 137 \times 2$ zones, respectively.}
\label{fig:POTENTIAL}
\end{figure*}

A snapshot of the resulting density distribution obtained with the
Yin-Yang grid is displayed in Fig.\,\ref{fig:RTCON} at epoch
$t=2.85\,$s. The left panel shows color coded contour lines in 3D, and
the right one a meridional cut at $\phi^{(n)}=0$. The contour lines are
drawn using different color tables for the Yin and Yang grid,
respectively.  Four distinct bubbles of rising low density fluid (Yin:
blue; Yang: bright gray) are clearly visible that reflect the initial
perturbation mode $(l,m)=(3,2)$. High density fluid (Yin: yellow/red;
Yang: dark gray/black) sinks down and settles at the inner part of the
sphere. One can also notice Kelvin-Helmholtz instabilities developing
at the surface of the bubbles. This is particularly obvious in the
meridional cut (right panel).  One of the RTI bubbles is within the
Yang grid, while the three others reside on the Yin grid. It is
obvious that the bubbles are distributed symmetrically following the
perturbation pattern regardless of the grid patch.  The 2D contour
lines shown in the right panel of Fig.\,\ref{fig:RTCON} emphasize this
fact.

The RTI bubbles grow with nearly the same growth rate in all four
(perturbation) directions, as can also be seen from Fig.\,
\ref{fig:RTGROWTH} that displays the position of each bubble's head
versus time. The four curves lie exactly on top of each other during
the phase of linear growth. There are slight discrepancies between the
four curves in the non-linear regime, because the linear grid
resolution in angular direction is slightly non-equidistant (due to
its $\theta$ dependence). Two curves from the Yin grid coincide
perfectly since they represent the two bubbles that lie symmetrically
above and below the equator in the Yin grid. The results confirm that
the Yin-Yang grid does not favor any angular direction on the
sphere. Since our aim was only to demonstrate this important fact, we
do not further analyze the growth rate of the RTI.

\begin{table*}
\begin{tabular}{|c|c|c|c|c|}
\hline
  \multirow{2}{*}{component} 
& \multicolumn{2}{|c|}{prolate spheroid} 
& \multicolumn{2}{|c|}{sphere} 
\\  
\cline{2-5}
 & Poisson solver & extended Poisson solver & 
   Poisson solver & extended Poisson solver \\
\hline 
 $\boldsymbol{\hat{r}}$ & $4.821\times 10^{-4}$ 
                        & $4.698\times 10^{-4}$ 
                        & $1.598\times 10^{-2}$
                        & $1.557\times 10^{-2}$ \\
 $\boldsymbol{\hat{\theta}}$ & $6.134\times 10^{-2}$ 
                             & $2.592\times 10^{-2}$ 
                             & $1.67 \times 10^{-2}$
                             & $1.67 \times 10^{-2}$ \\   
 $\boldsymbol{\hat{\phi}}$ & $1.245\times 10^{-2}$ 
                           & $2.435\times 10^{-3}$ 
                           & $1.655\times 10^{-2}$ 
                           & $1.655\times 10^{-2}$ \\                 
\hline 
\end{tabular}
\caption{Mean errors in the gravitational acceleration.}
\label{tab:dphicompa}
\end{table*}

\subsection{Gravitational Potential of Homogeneous Spheroids}
We investigate the accuracy of our gravity solver by calculating the
gravitational potential of homogeneous spheroids. We consider two
homogeneous self-gravitating configurations: a prolate spheroid with
an axis ratio of $0.7$, and a sphere. The configurations have a
constant density $\rho = 1\,$g/cm$^3$, and are embedded into a
homogeneous background of much lower density $\rho_b =
10^{-20}\,$g/cm$^3$ in order to minimize the background's contribution
to the gravitational potential. The semi-major axis of the spheroid
aligns with the $x$-axis, while its center is placed at the origin of
the Yin-Yang grid. To provide a more difficult test for our multipole
based gravity solver, we shift the center of the sphere off the origin
of the computational grid by more than one sphere radius.

The analytical form of the gravitational potential for both type of
configurations are known. The solution for the prolate spheroid can be
found in chapter\,3 of \citet{potential}, and the sphere's potential
can be easily calculated. Fig.\,\ref{fig:POTENTIAL} shows contour
lines of the gravitational potential for both cases in the meridional
plane $\phi^{(n)}=0$. The potential is calculated on a grid of $400
\times 92 \times 272 \times 2$ zones with $L=15$, where $L$ is the
number of spherical harmonics taken into account (see
section\,\ref{sec.gravpot}). The contour lines are smooth across the
Yin-Yang boundary for both the prolate spheroid and the sphere.

Concerning the convergence behavior of the solver, we consider various
grid resolutions and a number of spherical harmonics ranging up to
$L=25$ for this convergence test. The grid resolutions used in the
test are $400 \times 92 \times 272 \times 2$ zones, $400 \times 47
\times 137 \times 2$ zones, $200 \times 92 \times 272 \times 2$ zones,
and $200 \times 47 \times 137 \times 2$ zones, respectively. The
maximum and mean error of the gravitational potential are given as a
function of $L$ for both considered configurations in the middle and
right panels of Fig. \ref{fig:POTENTIAL}, respectively. Both errors
show a convergence behavior with higher grid resolution, and tend to
saturate at large values of $L$. This behavior is similar to what is
described in \citet{grav-solver}. In addition, for lower grid
resolution the accuracy saturates at a lower number of spherical
harmonics compared to calculations with a higher grid resolution. This
is expected since higher order terms in the multipole expansion are
not well represented on grids of lower angular resolution.

We also tested our extended Poisson solver discussed in
section\,\ref{sec.gravpot}.  In Table\,\ref{tab:dphicompa} we compare
the mean errors in the components of the gravitational acceleration
for both the prolate spheroid and the sphere test case computed with
the numerically differentiated gravitational potential given in
Eq.\,(\ref{eq:pot}) with those obtained from the analytic expression
given in Eqs.\,(\ref{eq:accel_r}), (\ref{eq:accel_theta}), and
(\ref{eq:accel_phi}), respectively.  We used a grid of $400 \times 92
\times 272 \times 2$ zones and $L=15$ for this comparison.

For the prolate spheroid test case the ``analytically'' obtained
accelerations exhibit a smaller mean error, especially for the
$\theta$- and $\phi$-component of the gravitational acceleration. This
results from a strong decrease of the maximum error, which is large in
regions where the angular components of the gravitational acceleration
approach zero, \ie near the major and minor axes of the prolate
spheroid. However, in these regions the accelerations in $\theta$ and
$\phi$-direction are orders of magnitude smaller than the radial
component. Thus, they contribute only a tiny fraction to the total
acceleration.  In the sphere test case both variants of the extended
Poisson solver produce similar mean errors. Based on these results we
conclude that the extended Poisson solver, which provides the
gravitational acceleration using analytic expressions, works
properly. Moreover, it gives a slightly more accurate gravitational
acceleration, as it does not involve numerically differencing the
gravitational potential. Nevertheless, for the reasons stated in
section\,\ref{sec.gravpot}, we prefer to use the Poisson solver of
\citet{grav-solver} in our simulations.

\subsection{Self-gravitating Polytropes}
Using our Yin-Yang grid based hydro-code we have also considered
self-gravitating, non-rotating and rotating equilibrium polytropes.
Both kinds of polytropes provide another test of the Poisson solver,
and a test of how well our hydrodynamics code can keep a
self-gravitating configuration in hydrostatic and stationary
equilibrium, respectively. In addition, the rotating polytrope also
serves to test the proper working of the Yin-Yang boundary treatment,
as it involves a considerable and systematic flow of mass, momentum
and energy flux across that boundary due to the polytrope's rotation.

The polytropes have a polytropic index $n=1$, a polytropic constant
$\kappa=1.455\times 10^5$, and a central density of $\rho_c =
7.905\times 10^{14}\,$g/cm$^3$.  For our test runs we interpolated
equilibrium polytropes calculated with the method of \citet{polytrope}
onto a Yin-Yang sphere, and simulated their dynamic evolution
(occurring as the interpolated configuration is not in perfect
hydrostatic equilibrium). The central region ($r < 1\,$km) of the
polytrope is cut out and replaced by a corresponding point mass to
allow for a larger time step.

We use an artificial atmosphere technique to handle those regions of
the computational grid that lie outside the (rotating, i.e.,
non-spherical) polytrope. The density in the atmosphere is set equal
to a value $\rho_{atm} = 10^{-10} \rho_{c}$, where $\rho_{c}$ is the
central density of the polytrope. Here, atmosphere denotes any grid
zone whose density is less than the cut-off density $\rho_{cut-off} =
10^{-7} \rho_{max}$.  Furthermore, for all zones in the atmosphere the
velocity is set to zero in order to keep the atmosphere quiet. This
procedure is applied at the end of every time step throughout the
simulation. A zero-gradient boundary condition is imposed at the outer
radial boundary, and a reflecting boundary condition at the inner
one. The polytrope's evolution is followed for $10\,$ms corresponding
to approximately $10$ dynamic time scales in order to check how well
the initial approximate equilibrium configuration is maintained by the
Yin-Yang code.

\begin{figure}
\includegraphics[width=0.5\textwidth]{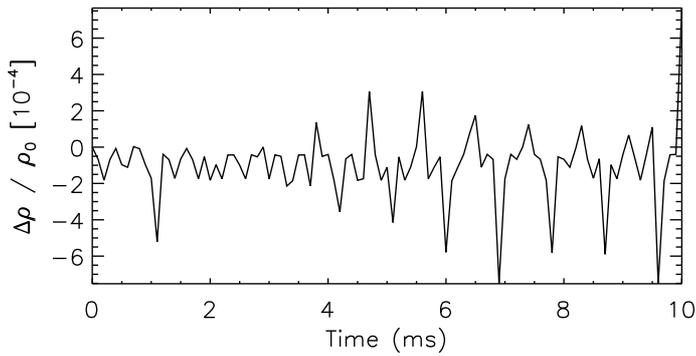}
\caption{Relative change of the central density of a non-rotating
  (nearly) equilibrium polytrope as a function of time.}
\label{fig:NONCENDEN}
\end{figure}

\begin{figure}
\includegraphics[width=0.5\textwidth]{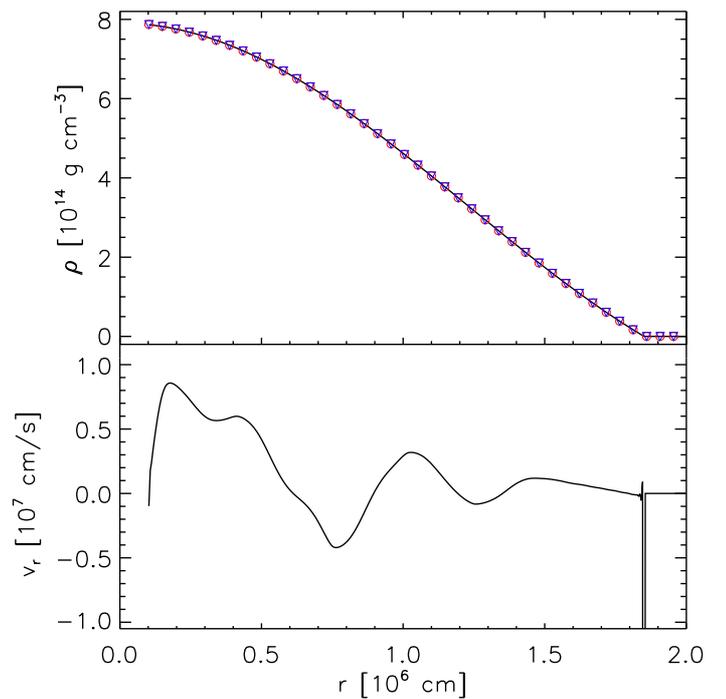}
\caption{Density (top) and radial velocity (bottom) of a non-rotating
  $n=1$ equilibrium polytrope as a function of radius after $t=10\,$ms
  of ``evolution''. In the top panel, the solid line shows the initial
  density profile. Red circles and blue triangles correspond to data
  from the Yin and the Yang grid, respectively.}
\label{fig:NONPROF}
\end{figure}

For the non-rotating polytrope, we employ a grid of $400 \times 20
\times 56 \times 2$ zones. Note that we are able to use a relatively
low angular resolution compared to the other tests, because the
problem has spherical symmetry.  Our results show that the polytrope
stays perfectly spherically symmetric throughout the simulation, and
that the non-radial velocities inside the polytrope remain zero. This
demonstrates that the Yin-Yang grid is able to preserve the initial
spherical symmetry.  Fig.\,\ref{fig:NONCENDEN} shows the evolution of
the central density (more precisely of the density of the innermost
radial zone at $r=1\,$km), which exhibits oscillations with an
amplitude of the order of $10^{-4}$ without any sign of a systematic
trend. Comparing the initial radial distributions of the density
(Fig.\,\ref{fig:NONPROF}, upper panel) and the radial velocity
(Fig.\,\ref{fig:NONPROF}, lower panel) of the polytrope with those
after $10\,$ms of evolution, we find no significant
deviations. Relative changes in the density profile are of the order
of $10^{-4}$, comparable to the size of the fluctuations of the
central density. Only for zones near the edge of the polytrope the
deviations can reach a level of up to $20\%$, in particular in the
zone next to the atmosphere. The figure also shows that data points
from the Yin and the Yang grid lie on top of each other confirming
that the code preserves the initial spherical symmetry of the
polytrope very well. Except for the zones at the polytrope's surface,
where the radial velocity is fluctuating at a level of approximately
$2 \times 10^8\,$cm/s, the radial velocities are less than
$10^6\,$cm/s (i.e., less than $0.1\%$ of the local sound speed). Thus,
we conclude that a non-rotating ($n=1$) equilibrium polytrope is
correctly handled by our Yin-Yang hydro-code.

\begin{figure}
\includegraphics[width=0.5\textwidth]{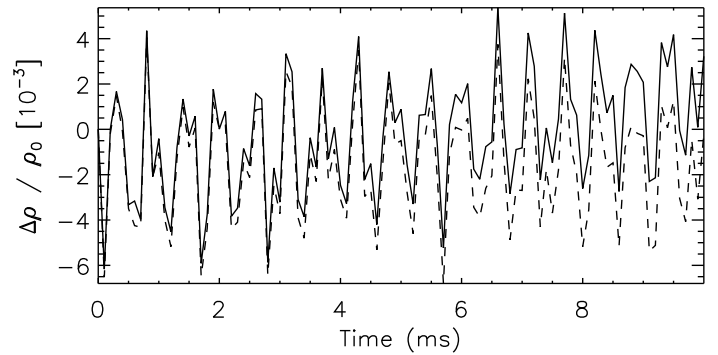}
\caption{Same as Fig.\,\ref{fig:NONCENDEN} but for a rotating
  polytrope. The solid and dashed curves show the relative variation
  of the density along an equatorial ray ($\theta^{(n)}=\pi/2$;
  $\phi^{(n)}=0$) and along the pole ($\theta^{(n)}=0$), respectively.}
\label{fig:ROTCENDEN}
\end{figure}

The rotating polytrope needs a higher grid resolution in
$\theta$-direction, as it is no longer spherically symmetric. Thus, we
used a grid resolution of $400 \times 92 \times 272 \times 2$ zones
for this simulation. The initial oblate equilibrium configuration has
an axis ratio of $0.7$. We, again, evolve the configuration for
$10\,$ms to test the correct treatment of the situation by our
Yin-Yang hydro-code.

\begin{figure}
\includegraphics[width=0.5\textwidth]{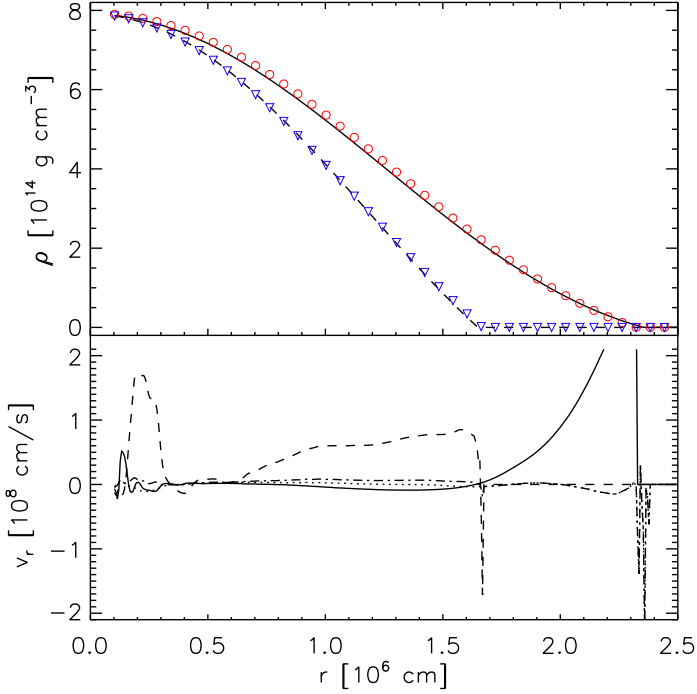}
\caption{Density (upper panel) and radial velocity (lower panel) of a
  $n=1$ rotating polytrope in stationary equilibrium as a function of
  radius after $t=10\,$ms of ``evolution''. In both panels the solid
  and dashed lines show the profiles along an equatorial ray
  ($\theta^{(n)}=\pi/2$, $\phi^{(n)}=0$) and along the pole
  ($\theta^{(n)}=0$), respectively.  Red circles and blue triangles in
  the upper panel correspond to data from the Yin and the Yang grid,
  respectively. In the lower panel, we show in addition time averaged
  (over the interval $t= [9, 10]\,$ms) velocity profiles along the
  equatorial ray (dotted) and the pole (dashed-dotted). }
\label{fig:ROTPROF}
\end{figure}

Fig. \ref{fig:ROTCENDEN} shows the relative variation of the central
density as a function of time along an equatorial ray
($\theta^{(n)}=\pi/2$; $\phi^{(n)}=0$) and along the pole ($\theta^{(n)}=0$),
respectively. One also recognizes a slight systematic trend in the
behavior of the density fluctuation, which is steeper along the
equator than at the pole. However, in both cases the relative increase
of the central density is very small ($\sim 10^{-3}$). The initial
radial density profiles along the pole and the equator do not show any
significant change during the $10\,$ms of evolution we have simulated
with the Yin-Yang code (Fig.\,\ref{fig:ROTPROF}, upper panel). The
axis ratio has slightly increased to a value of $0.719$. The radial
velocities (Fig. \ref{fig:ROTPROF}, lower panel) are larger than in
the non-rotating case by about an order of magnitude, because it is
obviously more difficult to keep a rotating polytrope in equilibrium
than a non-rotating (spherically symmetric) one. We again find the
largest radial velocities (a few times $10^8\,$cm/s) near the surface
of the polytrope, especially along the equator. However, these
velocities vary with time. When averaged over time (in the time
interval $t= [9, 10]\,$ms) the profiles become flatter and the
velocities smaller. This confirms that the polytrope is oscillating
around its equilibrium configuration.
          
\section{Conservation problem}
The Yin-Yang grid has a disadvantage common with other types of
overlapping grids \citep[see, e.g.,][]{chesshire_henshaw, wang,
  wu_etal}. The communication via interpolation between the two grid
patches does not guarantee conservation of conserved quantities even
though the finite-volume difference scheme employed on each grid patch
is conservative.  Non-conservation occurs when a flow across the
Yin-Yang boundary is present. This is the case in most of our tests
except for the simulation of the non-rotating polytrope that involves
only radial flow.

\begin{figure}
\includegraphics[width=0.5\textwidth]{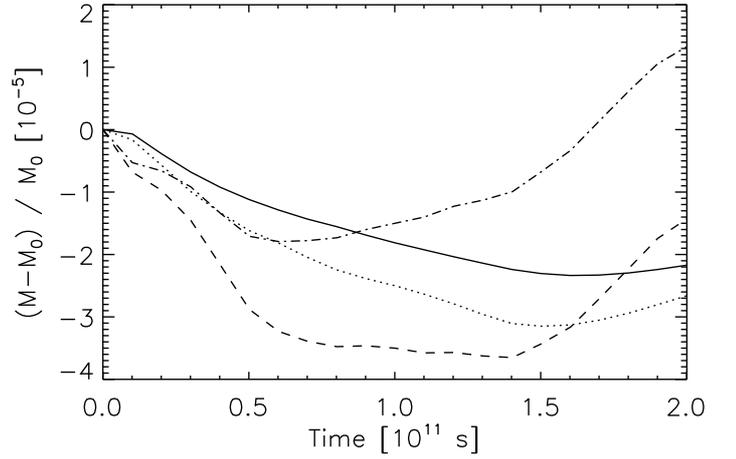}
\caption{Evolution of the relative mass loss, $(M-M_0)/M_0$, where
  $M_0$ is the initial total mass, for the Taylor-Sedov test simulated
  on three different (angular) grids with $400 \times 32 \times 92
  \times 2$ zones (\ie $3^\circ$ angular resolution; dashed line),
  $400 \times 92 \times 272 \times 2$ zones (\ie $1^\circ$ angular
  resolution; dotted line), and $400 \times 182 \times 542 \times 2$
  zones (\ie $0.5^\circ$ angular resolution; solid line),
  respectively.}
\label{fig:SEDCONSV}
\end{figure}

Nevertheless, we are still able to obtain sufficiently good results
for all the test simulations discussed in the previous section. The
degree of non-conservation is highly problem dependent. A simulation
involving a considerable and systematic flow across the Yin-Yang
boundary, as \eg in the case of the rotating polytrope, will result in
a larger degree of non-conservation. We observe that the total mass
increases by $0.07\%$ within $10\,$ms (or about ten dynamical
timescales) in the case of the rotating polytrope. For the
Taylor-Sedov test case, which is the cleanest test case in this
respect (as it involves, \eg no boundary effects like the shock tube,
and \eg no artificial atmosphere like the rotating polytrope), we find
a mass loss of the order of $10^{-5}$, only.  As
Fig.\,\ref{fig:SEDCONSV} demonstrates this mass loss can be reduced by
using a higher angular resolution.

Conservation of conserved scalar quantities can be obtained to machine
precision by applying the algorithm described in detail in
\citet{conserv}, and summarized below. According to this algorithm
scalar fluxes at the outer edges of boundary zones of both the Yin and
the Yang grid are replaced by scalar fluxes computed using only
``interior'' fluxes from adjacent grid zones.

\begin{figure}
\includegraphics[width=0.5\textwidth]{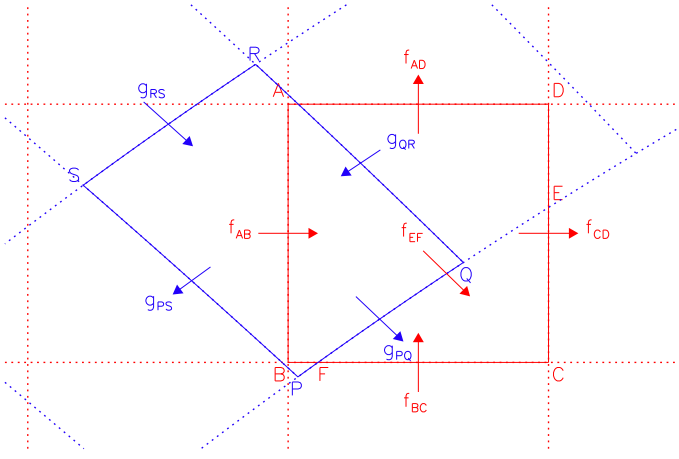}
\caption{Illustration of the Yin-Yang grid overlap configuration,
  where $PQRS$ is a grid zone at the boundary of the Yang grid (blue)
  which overlaps with the underlying grid zone $ABCD$ of the Yin grid
  (red).  Fluxes referring to the Yin and the Yang grid are denoted by
  $f$ and $g$, respectively. }
\label{fig:CONSV}
\end{figure}

As an illustration, consider the Yin-Yang grid overlap configuration
in Fig.\,\ref{fig:CONSV}, where $PQRS$ is a grid zone at the boundary
of the Yang grid (blue) which overlaps with the underlying grid zone
$ABCD$ of the Yin grid (red).  Fluxes referring to the Yin and the
Yang grid are denoted by $f$ and $g$, respectively.

The boundary flux $g_{PQ}$ of the Yang grid is replaced by the flux
\begin{equation}
 \label{eq:fluxpq}
 f_{PQ} = f_{FQ} + f_{PF}, 
\end{equation}
where $f_{FQ}$ and $f_{PF}$ are the fluxes through the segments
$\overline{FQ}$ and $\overline{PF}$, respectively.

The flux $f_{FQ}$ in Eq.(\ref{eq:fluxpq}) is calculated using
information from zone $ABCD$. The evolution of a scalar quantity
$\xi_{ABCD}$ of zone $ABCD$ is given by
\begin{equation}
 \label{eq:consv1}
 \xi_{ABCD}^{t+\Delta t} = \xi_{ABCD}^{t} + (f_{AB}-f_{CD}+f_{BC}-f_{AD}).
\end{equation}
Similarly, for the fraction of the zone $ABCD$ defined by the polygon
$ABFED$ one has,
\begin{equation}
 \label{eq:consv2}
 \xi_{ABFED}^{t+\Delta t} = \xi_{ABFED}^t +  
                          (f_{AB}-f_{CD}\frac{\overline{DE}}{\overline{CD}}
                          +f_{BC}\frac{\overline{BF}}{\overline{BC}}
                          -f_{AD}-f_{EF}).
\end{equation}
Assuming a piecewise constant state within the zone $ABCD$,
Eqs.(\ref{eq:consv1}) and (\ref{eq:consv2}) lead to
\begin{equation}
 \label{eq:consv3}
 \alpha(\xi_{ABCD}^{t+\Delta t}-\xi_{ABCD}^{t})=\xi_{ABFED}^{t+\Delta
        t}-\xi_{ABFED}^t
\end{equation}
where $\alpha$ is the overlapping volume fraction (area) described in
section\,3. Therefore,
\begin{equation}
 \label{eq:consv4}
 \alpha(f_{AB}-f_{CD}+f_{BC}-f_{AD}) = f_{AB} - 
                                      f_{CD} \frac{\overline{DE}}{\overline{CD}}
                                     +f_{BC}\frac{\overline{BF}}{\overline{BC}}
                                     -f_{AD}-f_{EF}
\end{equation}
Note that the flux $f_{EF}$ is the only unknown in
Eq.(\ref{eq:consv4}).  Since the intersection points $E$ and $F$ are
already known from the step to calculate the volume fraction $\alpha$,
the lengths of all segments can be obtained. The flux $f_{FQ}$ is then
given by
\begin{equation}
 \label{eq:consv5}
 f_{FQ} = f_{EF}\frac{\overline{FQ}}{\overline{EF}} \,.
\end{equation}
After obtaining the still missing flux $f_{PF}$ in
Eq.(\ref{eq:fluxpq}) by a similar procedure, the scalar quantity
$\xi_{PQRS}$ of the boundary zone $PQRS$ is updated according to
\begin{equation}
 \xi_{PQRS}^{t+\Delta t} = \xi_{PQRS}^{t} + (g_{QR}-g_{PS}+g_{RS}-f_{PQ}).
\end{equation}  
This procedure is then repeated to update all boundary grid zones.

After implementing the above algorithm we are able to conserve mass
and total energy up to machine precision. However, the conservation of
momentum is more complicated since the momentum equations in spherical
coordinates involve not only flux (\ie divergence) terms but also
source terms (due to the presence of fictitious and pressure forces),
and due to the ``mixing'' of momentum components as the Yin and Yang
grid patches are rotated relative to each other (see
Fig.\,\ref{fig:CONSV}).

As we have not yet devised and implemented a corresponding momentum
conservation algorithm, momentum is not yet perfectly conserved in our
code. For that reason we also refrain from using the scalar
conservation algorithm described above, since in some simulations (\eg
in the Taylor-Sedov explosion simulation) we encountered a negative
internal energy in some zones due to the inconsistency arising from
the perfect conservation of mass and total energy on one hand and the
imperfect conservation of momentum on the other hand.  In our test
runs the momentum violation is small, \eg amounting to 0.24\% (0.03\%)
angular momentum loss in the case of the rotating polytrope for a grid
with three (one) degree angular resolution.

\section{Performance and Efficiency}
One of the main purposes in implementing the Yin-Yang grid is to ease
the severe restriction imposed on the size of the time step for any
explicit hydrodynamics scheme by the CFL condition in the polar
regions of 3D simulations using a grid in spherical polar
coordinates. In most applications the size of the time step is
restricted most strongly by the size of the zones in $\phi$-direction,
which is smaller than the size in $\theta$-direction by the factor
$\sin \theta$ assuming an equal angular resolution $\delta \equiv
\Delta \theta = \Delta \phi$ in both angular directions.

For a spherical polar grid the factor $\sin \theta$ implies (assuming
zone centered variables) a minimum zone size
$$d^{sph}_{\phi} \equiv \delta\, \sin( \delta/2)$$
(in radians) in $\phi$-direction for the first zone next to the
pole. Typically, $\sin(\delta/2) \approx 10^{-2}$.  On the other hand,
applying the Yin-Yang grid yields
$$d^{YY}_{\phi} \equiv \delta\, \sin(\pi/4 - \delta/2)$$
for the size of the smallest zone in $\phi$-direction, which is
typically about $0.7$.  Hence, for the Yin-Yang grid the smallest zone
size in azimuthal direction is larger by the ratio
\begin{equation}
 \label{eq:gain}
  \frac{ d^{YY}_{\phi} }{ d^{sph}_{\phi} } = 
                 \frac{ \sin(\pi/4 - \delta /2) }{ \sin(\delta/2)} 
\end{equation}
compared to the spherical polar grid. 

Table\,\ref{tab:gain} gives the value of this ratio for grids of 
various angular resolution, and various computational domains. 
These numbers provide an estimate of the gain in computation time 
one can expect when using the Yin-Yang grid instead of the spherical 
polar grid.

\begin{table}
\begin{tabular}{|l|c|c|}
\hline
 computational domain            & angular grid resolution & gain factor \\  
\hline 
 full $4\pi$ sphere              & $3^\circ$               & 26 \\
                                 & $2^\circ$               & 40 \\
                                 & $1^\circ$               & 80 \\
\hline 
 sphere except for a cone         &                       &    \\
 of $5^\circ$ half opening angle   & $1^\circ$              &  7 \\
 cut-out at both poles            &                       &    \\
\hline
\end{tabular}
\caption{Expected gain factor when using the Yin-Yang grid. }
\label{tab:gain}
\end{table}

However, the gain factor calculated from the relative grid spacings
does not determine the gain in the size of the time step, as the latter
is given in a more complicated way by the CFL condition
\begin{eqnarray}
\label{CFL}
  \Delta t_{CFL} < C\, \Biggm(
                  \left| \frac{v_r     }{\Delta r}             \right| 
                  &+&
                  \left| \frac{v_\theta}{r\Delta\theta}         \right| +
                  \left| \frac{v_\phi  }{r\sin\theta\Delta\phi}
                  \right| 
\nonumber\\
                  &+& \sqrt{ \frac{c_s^2}{\Delta r^2 +
                                     (r\Delta\theta)^2 +
                                     (r\sin\theta\Delta\phi)^2 } }
                  \Biggm)^{-1/2} \, , 
\end{eqnarray}
where $C$, $v_r$, $v_\theta$, $v_\phi$, and $c_s$ are the Courant
factor, the flow velocities in radial, colatitude and azimuthal
direction, and the local sound speed, respectively. The CFL condition
shows that the increase in the size of the CFL time step is somewhat
smaller than implied by the gain factor resulting from the ratio of
the sizes of the smallest zones of the Yin-Yang grid and the spherical
polar grid. In addition, the increase of the time step is problem
dependent.

Besides the performance gain due to the increased size of the CFL time
step, the Yin-Yang grid also requires less computational zones to
cover the full sphere, and thus less computational time. For an
angular resolution $\delta$ the spherical polar grid needs
$$ (\pi/\delta) \times (2\pi/\delta)$$ 
zones to cover the full sphere, while the Yin-Yang grid requires only
$$ (\pi/2\delta+2) \times (3\pi/2\delta+2) \times 2 $$
zones. Hence, up to 25\% fewer computational zones are required. The
gain depends only weakly on angular resolution and is problem
independent.

However, employing the Yin-Yang grid also requires some extra amount
of computation compared to the spherical polar grid (see Sec.\,3). In
the following we only consider the extra costs of calculations during
the actual simulation, but not the extra costs arising during the
initialization, since these are negligible. We emphasize again that
there are two major extra sets of calculations necessary when applying
the Yin-Yang grid. The first set concerns the interpolation of the
ghost zone values that are needed for the communication between the
Yin and Yang grid patches. The second set arises from the
interpolation of the density onto the auxiliary spherical polar grid
grid and the interpolation of the gravitational potential back from
the auxiliary grid onto the Yin-Yang grid. Exploiting the algorithms
described in Sec.\,3, the computational cost for both parts is almost
negligible compared to the total computing time. Interpolation of the
ghost zone values requires only $2.3\%$ of the total computing time
per cycle in simulations with self-gravity, while the interpolation of
density and gravitational potential performed within the gravity
solver accounts for $1.5\%$ of the computing time needed for the
gravity solver.  This corresponds to approximately $0.3\%$ of the
computing time per cycle.

To obtain actual numbers for the gain, we performed several timing
tests including simulations with and without self-gravity using four
different grid resolutions. The tests were carried on an IBM Power6
using a single processor. According to these tests the computing time
per cycle for the Yin-Yang grid averaged over five cycles is
approximately $15\%$ and $20\%$ smaller than for the spherical polar
grid for simulations without self-gravity and with $2^\circ$ and
$1^\circ$ angular resolution, respectively. For simulations including
self-gravity, the gain factor decreases by $3\%$ approximately.

Concerning the gain from the less restrictive CFL condition, we
consider the case of the rotating polytrope since the size of the time
step does not vary much throughout the simulation. For an angular
resolution of $1^\circ$, we find a gain of approximately a factor of
63 when using the same Courant number both for the Yin-Yang grid and
the spherical polar grid.

\section{Conclusion}

A two-patch overset grid in spherical coordinates called the
``Yin-Yang'' grid is successfully implemented into our 3D Eulerian
explicit hydrodynamics code, PROMETHEUS, including in particular the
treatment of self-gravitating flows.  The Yin-Yang grid eases the
severe restriction of the time step size in the polar regions of the
sphere, because each Yin-Yang grid patch contains only the
low-latitude part of the usual spherical polar grid. From our
experiences, the implementation steps are easy and straightforward for
a hydrodynamics code using directional splitting and having 3D
spherical polar coordinates already implemented due to the simplicity
of the Yin-Yang transformation and its symmetry property.  Basically
it involves doubling the state variable arrays, calling the 1D core
hydrodynamics solver in angular directions for both the Yin and the
Yang grid, and adding a subroutine that handles the Yin-Yang
transformation and the interpolation of variables between both grids.

We validated the code with several standard hydrodynamic tests.  The
test results show good agreement with analytic solutions if these are
available.  Furthermore, as demonstrated by three of our test problems
-- a planar shock crossing the Yin-Yang grid boundary, an off-center
Taylor-Sedov explosion involving mass, momentum (all three components)
and energy flux across the Yin-Yang grid boundary, and a polytrope
whose rotation leads to considerable and systematic mass, momentum
(only angular components) and energy flux across the Yin-Yang grid
boundary -- the Yin-Yang grid does not introduce any numerical
artifact at the internal Yin-Yang boundary.  The tests also confirm
that the numerical solutions obtained with the Yin-Yang grid do not
show any evidence of a preferred radial direction, as it eliminates
numerical axes artifacts which seriously flaw the flow near the
coordinate symmetry axis when using a spherical polar grid.  Besides
successfully simulating a Taylor-Sedov explosion and self-gravitating
(rotating and non-rotating) equilibrium polytropes the code has also
passed another astrophysically relevant test involving the growth of
Rayleigh-Taylor instabilities.  

Because the communication between the two grid patches involves
interpolation, flows across the Yin-Yang boundary cause some small
amount of non-conservation of conserved quantities.  However, even for
the (in this respect) severe test case of the rotating polytrope
involving large flows across the Yin-Yang boundary, we observe only a
small amount (less than one percent) of non-conservation.

The Yin-Yang grid offers a large gain in computing time arising from
two sources. Firstly, the number of computational zones needed is
reduced by $20\%$ approximately depending on the angular
resolution. This gain reduces the computing time per cycle and is
problem independent. Secondly, the size of the CFL time step is
considerably enhanced, because the polar regions with converging
meridional coordinate lines are not present in case of the Yin-Yang
grid.  The corresponding gain in time step size highly depends on the
problem simulated. The extra costs for interpolation between the two
grid patches and the interpolation performed in the gravity solver are
negligible compared to the gain in the time step size.

In conclusion, our implementation of the Yin-Yang grid into the
multi-dimensional hydrodynamics code PROMETHEUS brings about the
possibility to simulate three dimensional self-gravitating
hydrodynamic flows in spherical coordinates which, in most cases, have
been computationally inaccessible up to now due to the prohibitively
large computational costs.  With the possibility to add more physics
such as neutrino transport (work in progress), the new code version
can be used to carry out, \eg core collapse supernova simulations in
3D.

\begin{acknowledgements}
%
This research was supported by the Deutsche Forschungsgemeinschaft
through the Transregional Collaborative Research Centers SFB/TR~27
``Neutrinos and Beyond'' and SFB/TR\,7 ``Gravitational Wave
Astronomy'', and the Cluster of Excellence EXC\,153 ``Origin and
Structure of the Universe''. The simulations were performed at the
Rechenzentrum Garching (RZG) of the Max-Planck-Society.
We would like to thank Prof.\,A.\,Kageyama for sharing with us his
Yin-Yang interpolation routine for scalar/vector fields, and the
anonymous referee for his/her useful and supportive comments.
\end{acknowledgements}

\bibliographystyle{aa}
\bibliography{13435}
\end{document}